%% file: Main+Supp.tex
\documentclass[11pt]{article}

\input{packages-v2}

\usepackage{helvet}
\usepackage{graphicx,amssymb,amsmath,amsfonts,amsthm,color,url, placeins}
\usepackage{wrapfig}
\usepackage{tabstackengine}
\usepackage{multirow,ctable} 
\usepackage{makecell}

\renewcommand\section{\@startsection{section}{1}{\z@}%
                                   {-3.0ex \@plus -1ex \@minus -.2ex}%
                                   {1.5ex \@plus.2ex}%
                                   {\normalfont\sffamily\large\bfseries}}
\renewcommand\subsection{\@startsection{subsection}{2}{\z@}%
                                     {-2.75ex\@plus -1ex \@minus -.2ex}%
                                     {1.5ex \@plus .2ex}%
                                   {\normalfont\sffamily\large}}
\renewcommand\subsubsection{\@startsection{subsubsection}{3}{\z@}%
                                     {-2.75ex\@plus -1ex \@minus -.2ex}%
                                     {1.5ex \@plus .2ex}%
                                   {\normalfont\sffamily\large}}

\newcommand{\od}{\stackrel{\mbox {\tiny {def}}}{=}}

\def\RR{\mathbb{R}}

\def\d{\mathrm{d}}

\def\RR{\mathbb{R}}

\def\RR{\mathbb{R}}

\def\max{\mathrm{max}}

\def\supp{\operatorname{supp}}

\def\od{\stackrel{\mathrm{def}}{=}}

\def\supp{\operatorname{supp}}

\def\FP{\operatorname{FP}}

\usepackage{color}
\definecolor{cherry}{rgb}{0.9,.1,.2}

\begin{document}

\part*{Core motifs predict dynamic attractors in combinatorial threshold-linear networks}
\noindent Caitlyn Parmelee, Samantha Moore, Katherine Morrison*, Carina Curto*  \\

\section*{Abstract}
\noindent Combinatorial threshold-linear networks (CTLNs) are a special class of inhibition-dominated TLNs defined from directed graphs. Like more general TLNs, they display a wide variety of nonlinear dynamics including multistability, limit cycles, quasiperiodic attractors, and chaos. In prior work, we have developed a detailed mathematical theory relating stable and unstable fixed points of CTLNs to graph-theoretic properties of the underlying network. Here we find that a special type of fixed points, corresponding to {\it core motifs}, are predictive of both static and dynamic attractors. Moreover, the attractors can be found by choosing initial conditions that are small perturbations of these fixed points. This motivates us to hypothesize that dynamic attractors of a network correspond to unstable fixed points supported on core motifs. We tested this hypothesis on a large family of directed graphs of size n=5, and found remarkable agreement.  Furthermore, we discovered that core motifs with similar embeddings give rise to nearly identical attractors. This allowed us to classify attractors based on structurally-defined graph families. Our results suggest that graphical properties of the connectivity can be used to predict a network's complex repertoire of nonlinear dynamics.

\section*{Introduction}

The vast majority of the literature on attractor neural networks has focused on fixed point attractors. The typical scenario is that of a network that contains either a discrete set of stable fixed points, as in the Hopfield model, or a continuum of marginally stable fixed points, as in continuous attractor networks. Depending on initial conditions, or in response to external inputs, the activity of the network converges to one of these fixed points. These are sometimes referred to as {\it static} attractors, because the fixed point is in an equilibrium or steady state. But neural networks, even very simple ones like threshold-linear networks (TLNs), can also exhibit {\it dynamic} attractors with periodic, quasiperiodic, or even chaotic orbits. What network architectures support these more complex attractors? And can their existence be predicted based on the structure of the underlying connectivity graph?

In this work, we tackle these questions in the context of combinatorial threshold-linear networks (CTLNs), which are a special class of TLNs defined from directed graphs. TLNs have been widely used in computational neuroscience as a framework for modeling recurrent neural networks, including associative memory networks \cite{AppendixE,Seung-Nature,HahnSeungSlotine, XieHahnSeung, flex-memory, net-encoding, pattern-completion, Fitzgerald2020}. And CTLNs are a subclass that are especially tractable mathematically \cite{CTLN-preprint, book-chapter, fp-paper, stable-fp-paper}.
What graph structures support dynamic attractors in CTLNs? 

We begin by observing an apparent correspondence between a network's {\it minimal} fixed points and its attractors. This leads us to hypothesize that fixed points supported on a special class of subgraphs, called {\it core motifs}, are predictive of both static and dynamic attractors. Next, we test this hypothesis on a large family of CTLNs on small graphs of $n=5$ nodes. We find that, with few exceptions, fixed points supported on core motifs correspond precisely to the attractors of the network. Moreover, we find that core motifs with similar embeddings in the larger network give rise to nearly identical attractors. This enables us to classify the observed attractors and identify common structural properties of the graphs that support them. Our results illustrate how the structure of a network can be used to predict the emergence of various attractors by way of graphical analysis.

\section*{Methods and Models}
We study dynamic attractors in a family of threshold-linear networks (TLNs).
The firing rates $x_1(t),\ldots,x_n(t)$ of $n$ recurrently-connected neurons evolve in time according to the standard TLN equations:
\begin{equation}\label{eq:dynamics}
\dfrac{dx_i}{dt} = -x_i + {\left[\sum_{j=1}^n W_{ij}x_j+b_i \right]}_+, \quad i = 1,\ldots,n
\end{equation}
where $n$ is the number of neurons.
The dynamic variable $x_i(t) \in \RR_{\geq 0}$ is the activity level (or ``firing rate'') of the $i^\textrm{th}$ neuron, and $b_i$ can represent a threshold or an external input.  The values $W_{ij}$ are entries of an $n \times n$ matrix of real-valued connection strengths. The threshold nonlinearity ${[\cdot]}_+ \od \max\{0,\cdot\}$ is critical for the model to produce nonlinear dynamics; without it, the system would be linear.

\subsubsection*{CTLNs} Combinatorial threshold-linear networks (CTLNs) are a special case of inhibition-dominated TLNs, where we restrict to having only two values for the connection strengths $W_{ij}$. These are obtained as follows from a directed graph $G$, where $j \to i$ indicates that there is an edge from $j$ to $i$ and $j \not\to i$ indicates that there is no such edge:
\begin{equation} \label{eq:binary-synapse}
W_{ij} = \left\{\begin{array}{ll} \phantom{-}0 & \text{ if } i = j, \\ -1 + \varepsilon & \text{ if } j \rightarrow i \text{ in } G,\\ -1 -\delta & \text{ if } j \not\rightarrow i \text{ in } G. \end{array}\right. \quad \quad \quad \quad
\end{equation}
Additionally, CTLNs typically have a constant external input $b_i=\theta$ in order to ensure the dynamics are internally generated rather than inherited from a changing or spatially heterogeneous input.  A CTLN is thus completely specified by the choice of a graph $G$, together with three real parameters: $\varepsilon, \delta$, and $\theta$.  We additionally require that $\delta >0$, $\theta>0$, and $0 < \varepsilon < \dfrac{\delta}{\delta+1}$. When these conditions are met, we say the parameters are within the \emph{legal range}. (The upper bound on $\varepsilon$ ensures that subgraphs consisting of a single directed edge $i \to j$ are not allowed to support stable fixed points \cite{CTLN-preprint}.) Note that the upper bound on $\varepsilon$ implies $\varepsilon < 1$, and so the $W$ matrix is always effectively inhibitory. For fixed parameters, only the graph $G$ varies between networks. Unless otherwise noted, the simulations presented here have parameters $\theta=1$, $\varepsilon=0.25$, and $\delta=0.5$. We will refer to these as the {\it standard parameters}.

\begin{figure}[!h]
\begin{center}
\includegraphics[width=4.5in]{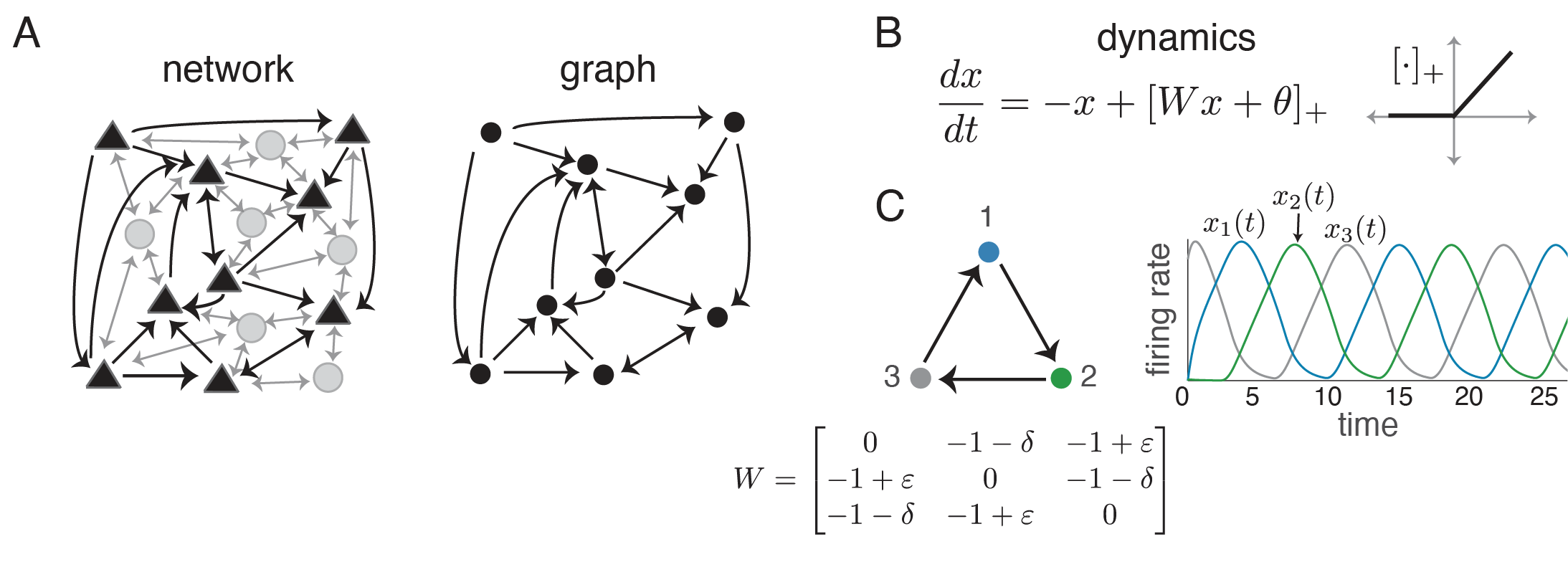}
\end{center}
\caption{{\bf CTLNs.} (A) A neural network with excitatory pyramidal neurons (triangles) and a background network of inhibitory interneurons (gray circles) that produces a global inhibition. The corresponding graph (right) retains only the excitatory neurons and their connections. (B) TLN dynamics and the graph of the threshold-nonlinearity ${[\cdot]}_+ = \max\{0,\cdot\}$. (C) A graph that is a 3-cycle (left) and its corresponding CTLN matrix $W$. (Right) A solution for the corresponding CTLN, with parameters $\varepsilon=0.25, \delta = 0.5$, and $\theta =1$, showing that network activity follows the arrows in the graph.  Peak activity occurs sequentially in the cyclic order 123.}
\label{fig:network-setup}
\end{figure}

We interpret the CTLN as modeling a network of $n$ excitatory neurons, whose net interactions are effectively inhibitory due to a strong global inhibition (Fig~\ref{fig:network-setup}A). When $j \not\to i$, we say $j$ \emph{strongly inhibits} $i$; when $j \to i$, we say $j$ \emph{weakly inhibits} $i$.  
Note that because $-1-\delta < -1 < -1+\varepsilon$, when $j \not\to i$ neuron $j$ inhibits $i$ \emph{more} than it inhibits itself via its leak term; when $j \to i$, neuron $j$ inhibits $i$ \emph{less} than it inhibits itself.  These differences in inhibition strength cause the activity to follow the arrows of the graph (see Fig~\ref{fig:network-setup}C).

\subsubsection*{Fixed points} Stable fixed points of a network are of obvious interest because they correspond to static attractors \cite{HahnSeungSlotine, net-encoding, pattern-completion, stable-fp-paper}. One of the most striking features of CTLNs, however, is the strong connection between unstable fixed points and dynamic attractors \cite{book-chapter}. This is our main focus here.

A fixed point $x^* \in \mathbb{R}^n$ of a TLN is a solution that satisfies $dx_i/dt|_{x=x^*} = 0$ for each $i \in [n]$, where $[n] \od \{1, \ldots, n\}$. The  {\it support} of a fixed point is the subset of active neurons, $\supp{x^*} = \{i \mid x^*_i>0\}$. CTLNs (and TLNs) are piecewise-linear dynamical systems, and we typically require a nondegeneracy condition that is generically satisfied and implies nondegeneracy in each linear regime \cite{fp-paper}. As a result, we have that for a given network there can be at most one fixed point per support.  Thus, we can label all the fixed points of a network by their support, $\sigma = \supp{x^*} \subseteq [n].$ We denote this collection of supports by:
\vspace{-.05in}
\[\FP(G) = \FP(G, \varepsilon, \delta)\od \{\sigma \subseteq [n] ~|~  \sigma \text{ is a fixed point support of the associated CTLN} \}.\]
Note that once we know $\sigma \in \FP(G)$ is the support of a fixed point, the fixed point itself is easily recovered. Outside the support, we must have $x_i^* = 0$ for all $i \not\in \sigma$. Within the support, $x^*$ is given by:
$$x_\sigma^* = \theta {(I-W_\sigma)}^{-1} 1_\sigma,$$
where $W_\sigma$ is the induced submatrix obtained by restricting rows and columns to $\sigma$, and  $1_\sigma$ is a column vector of all $1$s of length $|\sigma|$.

A useful fact is that a fixed point for a CTLN with graph $G$ is also a fixed point for any subnetwork containing its support. A {\it subnetwork} supported on $\sigma$ is a CTLN for the {\it induced subgraph} $G\vert_\sigma$ obtained from $G$ by restricting to the vertices of $\sigma$ and keeping only edges $i \to j$ for $i,j \in \sigma$ (see \cite{fp-paper}). In particular, if $\sigma \in \FP(G)$, then $\sigma \in \FP(G\vert_\sigma)$.  The converse is not true: one can have $\sigma \in \FP(G\vert_\sigma)$ in the subnetwork, but the fixed point may {\it not survive} the embedding into the larger network, and so $\sigma \notin \FP(G)$. 

\subsubsection*{Graph rules}  In prior work, a series of {\it graph rules} were proven that can be used to determine fixed points of a CTLN by analyzing the structure of the graph $G$ \cite{fp-paper, stable-fp-paper}.  These rules are all independent of the choice of parameters $\varepsilon, \delta$, and $\theta$. Some of the simplest graph rules are also quite powerful, and can be used to fully determine $\FP(G)$ for many graphs. Note that these are only valid for {\it nondegenerate} CTLNs, a condition defined in \cite{fp-paper} that generically holds.

\begin{table}[!h]
\centering
\begin{tabular}{c|l}
 {rule name} & {graph rule (theorem)}\\
 \hline
 \hline
sources & if $i \in \sigma$ is a proper source in $G\vert_\sigma$ or in $G$, then $\sigma \notin \FP(G)$\\
 \hline
sinks & if $\sigma = \{i\}$ is a singleton, then $\sigma \in \FP(G)$ iff $i$ is a sink of $G$\\
\hline
uniform & if $G\vert_\sigma$ is uniform in-degree $d$, then \\
in-degree & $\sigma \in \FP(G)\; \Leftrightarrow \; \forall\; k \notin \sigma$, $k$ receives at most $d$ edges from $\sigma$\\
 \hline
parity & the total number of fixed points, $|\FP(G)|$, is odd\\
\hline
\end{tabular}
\vspace{.15in}
\caption{\bf Graph rules for CTLNs.}
\label{table:graph-rules}
\end{table}
  
\begin{figure}[!h]
\begin{center}
\includegraphics[width=.95\textwidth]{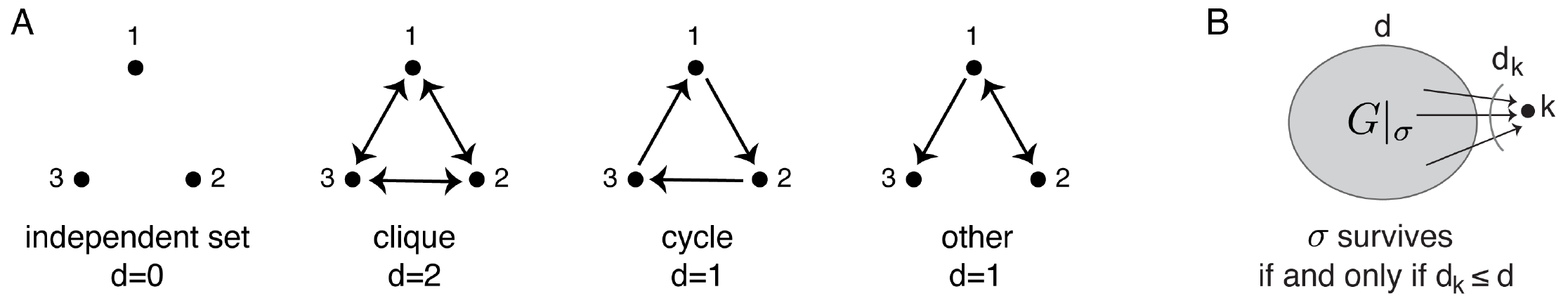}
\end{center}
\caption{{\bf Uniform in-degree graphs.} (A) All $n=3$ graphs with uniform in-degree. (B) Cartoon showing survival rule for an arbitrary subgraph with uniform in-degree $d$.}
\label{fig:ufd-examples}
\end{figure}

Table~\ref{table:graph-rules} lists a few of these rules, which were all proven in \cite{fp-paper}. To state the rules, we need some graph-theoretic terminology. A vertex $i$ of a graph $G$ is a {\it source} if it has no incoming edges $j \to i$, and it is a {\it proper source} if it also has at least one outgoing edge $i \to j$. A {\it sink} is a vertex with no outgoing edges.  A graph is {\it uniform in-degree} with degree $d$ if all vertices receive exactly $d$ incoming edges (see Fig~\ref{fig:ufd-examples}). Note that a vertex can be a source or a sink in an induced subgraph, $G\vert_\sigma$, but not in the full graph $G$. Similarly, a subgraph $G\vert_\sigma$ may be uniform in-degree even if the vertices of $\sigma$ do not have the same in-degree within the full graph $G$. 

The sources rule implies that there can be no fixed points supported on a pair of vertices $\{i,j\}$ corresponding to a single directed edge, $i \to j$, since $i$ is a proper source in the induced subgraph. The sinks rule tells us that the only singletons that can support fixed points correspond to sinks of $G$. The uniform in-degree rule implies that cycles, which have uniform in-degree $d=1$, support fixed points if and only if there is no vertex outside the cycle receiving two or more edges from it. It also implies that cliques of size $m$, which have uniform in-degree $m-1$, support fixed points if and only if they are {\it target-free}: that is, if and only if there is no vertex $k$ outside the clique receiving edges from all $m$ vertices of the clique. In fact, we have shown in \cite{CTLN-preprint} that target-free cliques, including sinks, correspond to {\it stable} fixed points. A third consequence of the uniform in-degree rule is that if $G\vert_\sigma$ is an independent set (uniform in-degree 0), then $\sigma \in \FP(G)$ if and only if each $i \in \sigma$ is a sink in $G$.

\begin{figure}[!ht]
\begin{center}
\includegraphics[width=\textwidth]{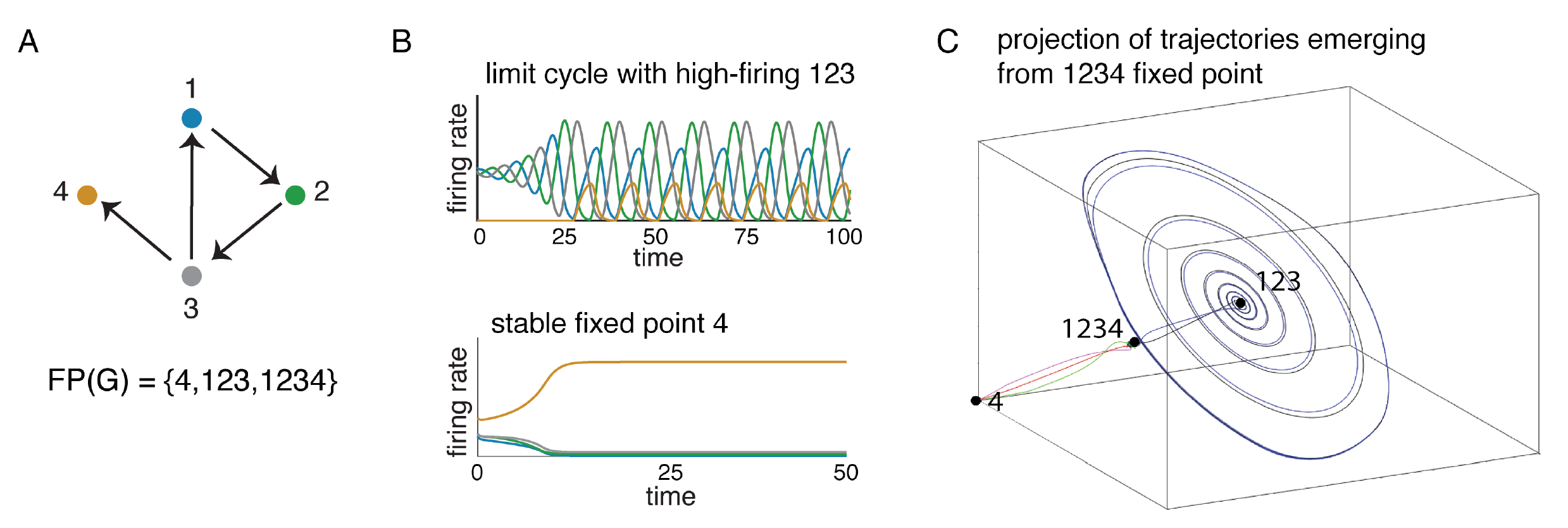}
\end{center}
\caption{{\bf An example CTLN and its attractors.} (A) The graph of a CTLN. Using graph rules, we can compute $\FP(G)$. (B) Solutions to the CTLN with the graph in panel A using the standard parameters $\theta=1$, 
$\varepsilon=0.25$, and $\delta=0.5$. (Top) The initial condition was chosen as a small perturbation of the fixed point supported on $123$. The activity quickly converges to a limit cycle where the high-firing neurons are the ones in the fixed point support. (Bottom) A different initial condition yields a solution that converges to the static attractor corresponding to the stable fixed point on node $4$. (C) The three fixed points are depicted in a three-dimensional projection of the four-dimensional state space. Perturbations of the fixed point supported on $1234$ produce solutions that either converge to the limit cycle shown in panel B, or to the stable fixed point. This fixed point thus lives on the boundary of the two basins of attraction, and behaves as a ``tipping point'' between the two attractors.}
\label{fig:tadpole}
\end{figure}

To see how these graph rules, together with parity, can be used to work out $\FP(G)$, consider the graph in Fig~\ref{fig:tadpole}A. From the sinks rule we immediately see that $\{4\} \in \FP(G)$, but no other singleton supports a fixed point. From the sources rules we see that none of the pairs $\{1,2\}$, $\{2,3\}$, $\{1,3\}$, and $\{3,4\}$ are in $\FP(G)$, as they correspond to edges having a proper source. Using the uniform in-degree rule we can also rule out the independent sets $\{1,4\}$ and $\{2,4\}$, since $1$ and $2$ are not sinks. On the other hand, the uniform in-degree rule does imply that the $3$-cycle $\{1,2,3\} \in \FP(G)$. All other subsets of size three can be ruled out because they have a proper source in the induced subgraph $G\vert_\sigma$. 

After checking all subsets of vertices of size $1$, $2$, or $3$, we have found that only $4, 123 \in \FP(G)$. Here we are simplifying notation for fixed point supports by writing $4$ instead of $\{4\}$ and $123$ instead of $\{1,2,3\}$. Now we can use parity to conclude that we must also have $1234 \in \FP(G)$, since this is the only remaining support and the total number of fixed points must be odd. Note that the rules we have used apply irrespective of the choice of $\varepsilon, \delta$ or $\theta$, provided they are in the legal range. It follows that $\FP(G) = \{4, 123, 1234\}$ is invariant under changes of these parameters.

\section*{Results}

\subsection*{Core fixed points and core motifs}

The starting point for this work was the following remarkable observation about CTLNs: namely, that their fixed points appear to give rise to both {\it static} and {\it dynamic} attractors. Consider again the network in Fig~\ref{fig:tadpole}A. From graph rules, we already saw that $\FP(G) = \{4, 123, 1234\}$, irrespective of the choice of $\varepsilon, \delta$ or $\theta$. The support $4$ corresponds to a sink in the graph, and gives rise to a stable fixed point (i.e., a static attractor). Initial conditions that are small perturbations of this fixed point will result in the network activity converging back to it (Fig~\ref{fig:tadpole}B, bottom). In contrast, the fixed point supported on $123$ gives rise to a limit cycle (i.e., a dynamic attractor). Initial conditions near this fixed point result in activity that converges to a periodic trajectory whose high-firing neurons are $1, 2$, and $3$ (Fig~\ref{fig:tadpole}B, top). The last fixed point, with full support $1234$, does not have a corresponding attractor. Initial conditions that are small perturbations of this fixed point can either converge to the attractor for $4$ or for $123$ (Fig~\ref{fig:tadpole}C). This fixed point lies on the boundaries of the two basins of attraction, and can thus be considered a ``tipping point.''

Which fixed points correspond to attractors, and which ones are tipping points? Fig~\ref{fig:tadpole} provides a hint: the fixed points giving rise to attractors have supports $4$ and $123$, which are {\it minimal} under inclusion in $\FP(G)$. The last fixed point, $1234$, has a support that contains smaller fixed point supports. We will call a fixed point {\it minimal} in $G$ if its support $\sigma$ is minimal in $\FP(G)$.

\begin{figure}[!ht]
\begin{center}
\includegraphics[width=.95\textwidth]{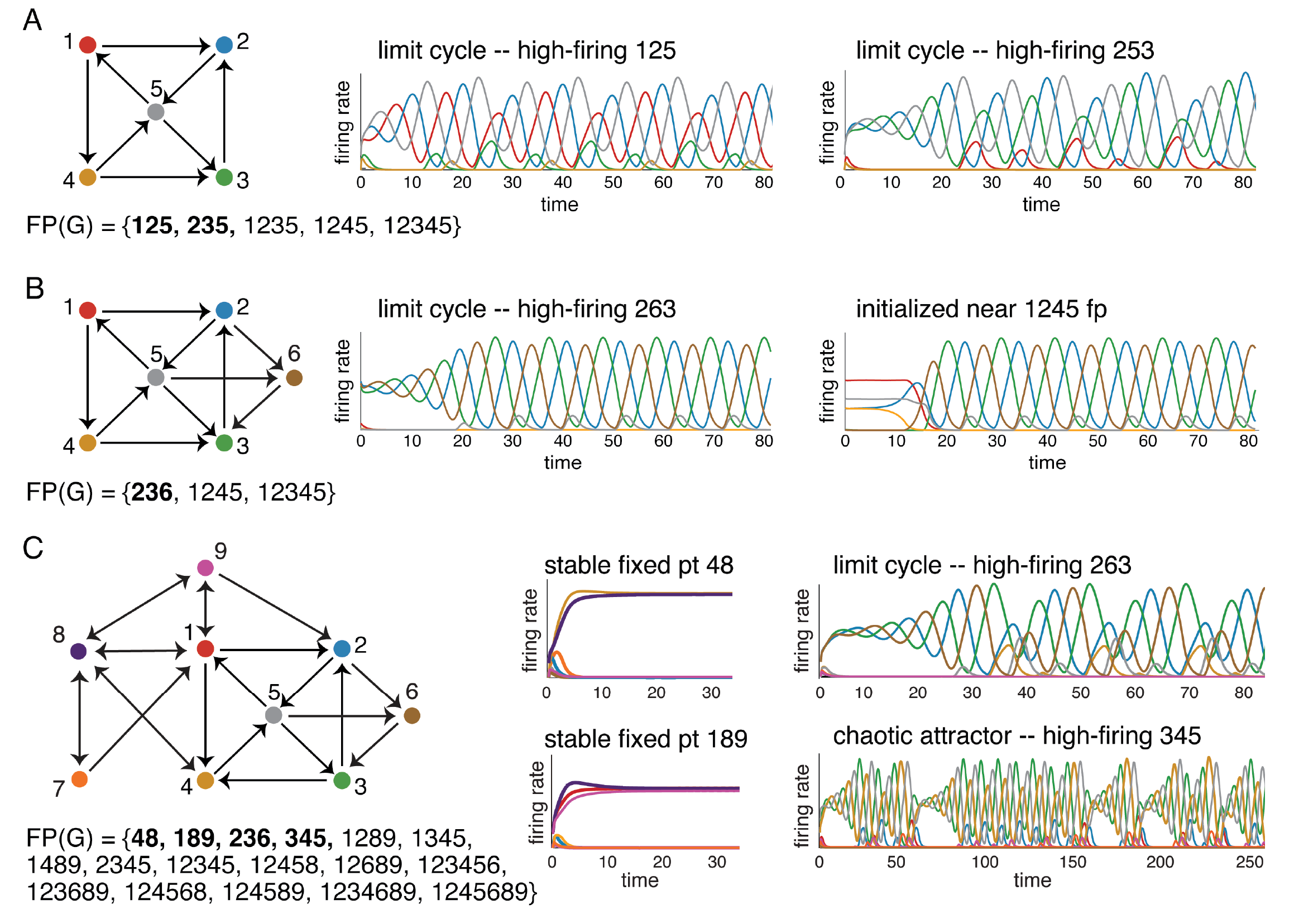}
\end{center}
\caption{{\bf Correspondence between core fixed points and attractors.} For each of the three graphs, $\FP(G)$ was computed using graph rules. Minimal fixed points that are also core fixed points are shown in bold. (A) A network on five nodes with two core fixed points supported on $125$ and $235$. Each of the two attractors of the network can be obtained via an initial condition that is a perturbation of one of these fixed points. The first attractor follows the cycle $125$ in the graph, while the second one follows the cycle $253$. (B) A network with the same graph as in A, except for the addition of node $6$. Although there are two minimal fixed points, supported on $236$ and $1245$, only the fixed point for $236$ is core and yields an attractor. Initial conditions near the $1245$ fixed point (denoted $1245$ fp) produce solutions that stay near the (unstable) fixed point for some time, but eventually converge to the same $236$ attractor.
(C) A larger network built by adding nodes $7, 8$, and $9$ to the graph in B, and flipping the $4 \to 3$ edge. This CTLN has four core fixed points, and no other minimal fixed points. Each core fixed point has a corresponding attractor: stable fixed points supported on $48$ and $189$, a limit cycle supported on $236$, and a chaotic attractor for $345$.}
\label{fig:rule-of-thumb}
\end{figure}

Next, consider Fig~\ref{fig:rule-of-thumb}A. Using the graph rules in Table~\ref{table:graph-rules}, together with two additional rules in~\cite{fp-paper} (Lemma 21 and graphical domination), we can work out $\FP(G)$ for this network as well. We find that there are two minimal fixed points, supported on $125$ and $235$. Consistent with our previous observations, each of these fixed points has a corresponding attractor. Specifically, we say that a fixed point {\it corresponds} to an attractor if 

\begin{itemize}
\item[(i)] initial conditions that are small perturbations from the fixed point lead to solutions that converge to the attractor, and
\item[(ii)] the high-firing neurons in the attractor match the support of the fixed point.
\end{itemize}

Just as we saw in Fig~\ref{fig:tadpole}, the induced subgraphs $G\vert_\sigma$ corresponding to the minimal fixed point supports $\sigma = 125$ and $\sigma = 235$ are $3$-cycles. Note, however, that the graph in Fig~\ref{fig:rule-of-thumb}A actually has a third $3$-cycle, $145$, yet this one does not support a fixed point of $G$ and also has no corresponding attractor. The reason $145 \notin \FP(G)$ is because node $3$ receives two edges from $145$, and so the uniform in-degree rule tells us the fixed point does not survive in the larger graph. In contrast, both $125$ and $235$ do have fixed points that survive to the full network. We see from this example that it is not enough to have a subgraph that supports an attractor. The $3$-cycle whose fixed point does not survive the embedding has no corresponding attractor.

Does every minimal fixed point of $G$ have a corresponding attractor? Unfortunately, minimality is not enough. Fig~\ref{fig:rule-of-thumb}B depicts a network built from the one in panel A by adding a single node, $6$. Because $6$ receives edges from both $2$ and $5$, it ensures that the $125$ and $235$ fixed points do not survive to the full network. As a result, $1235$ also dies but the fixed point $1245$ remains and becomes minimal. This fixed point does {\it not} have a corresponding attractor, however. Small perturbations of the fixed point for $1245$ lead the network to converge to the attractor corresponding to the other minimal fixed point, $236$. The main difference between these minimal fixed points is that $1245$ is not minimal in its own subnetwork, $G|_{1245}$, while $236$ is still minimal in $G|_{236}$.
This motivates the following definition:

\begin{definition}[core fixed point] We say that a fixed point of a CTLN on a graph $G$ is a {\it core fixed point} if its support $\sigma \in \FP(G)$ is minimal (under inclusion) in $\FP(G)$ and is also minimal in $\FP(G\vert_\sigma)$.
\end{definition}

 Equivalently, $\sigma$ is the support of a core fixed point if and only if $\sigma \in \FP(G)$ and 
 $\FP(G\vert_\sigma) = \{\sigma\}$. This is because the minimality of $\sigma$ in $G\vert_\sigma$ guarantees that $\sigma$ is the unique fixed point of $G\vert_\sigma$ and also that it is minimal in $G$ if it survives to $\sigma \in \FP(G)$. The converse, however, is not true. One can have $\sigma$ minimal in $G$ but not minimal in $G\vert_\sigma$, simply because the fixed points below it in the subnetwork did not survive to the larger one. This is what happened with the $\sigma = 1245$ fixed point in Fig~\ref{fig:rule-of-thumb}B. These are precisely the minimal fixed points we are ruling out with the above definition.

Since core fixed points must satisfy $\FP(G\vert_\sigma) = \{\sigma\}$, there are only certain subgraphs that can support them. It is useful to give these graphs their own name:

\begin{definition}[core motifs] Let $G\vert_\sigma$ be a subgraph of $G$ that satisfies $\FP(G\vert_\sigma) = \{\sigma\}$. Then we say that $G\vert_\sigma$ is a {\it core motif} of $G$. When the graph in question is understood, we may also refer to the support itself, $\sigma$, as a core motif.
\end{definition}

\begin{figure}[!ht]
\begin{center}
\includegraphics[width=5.75in]{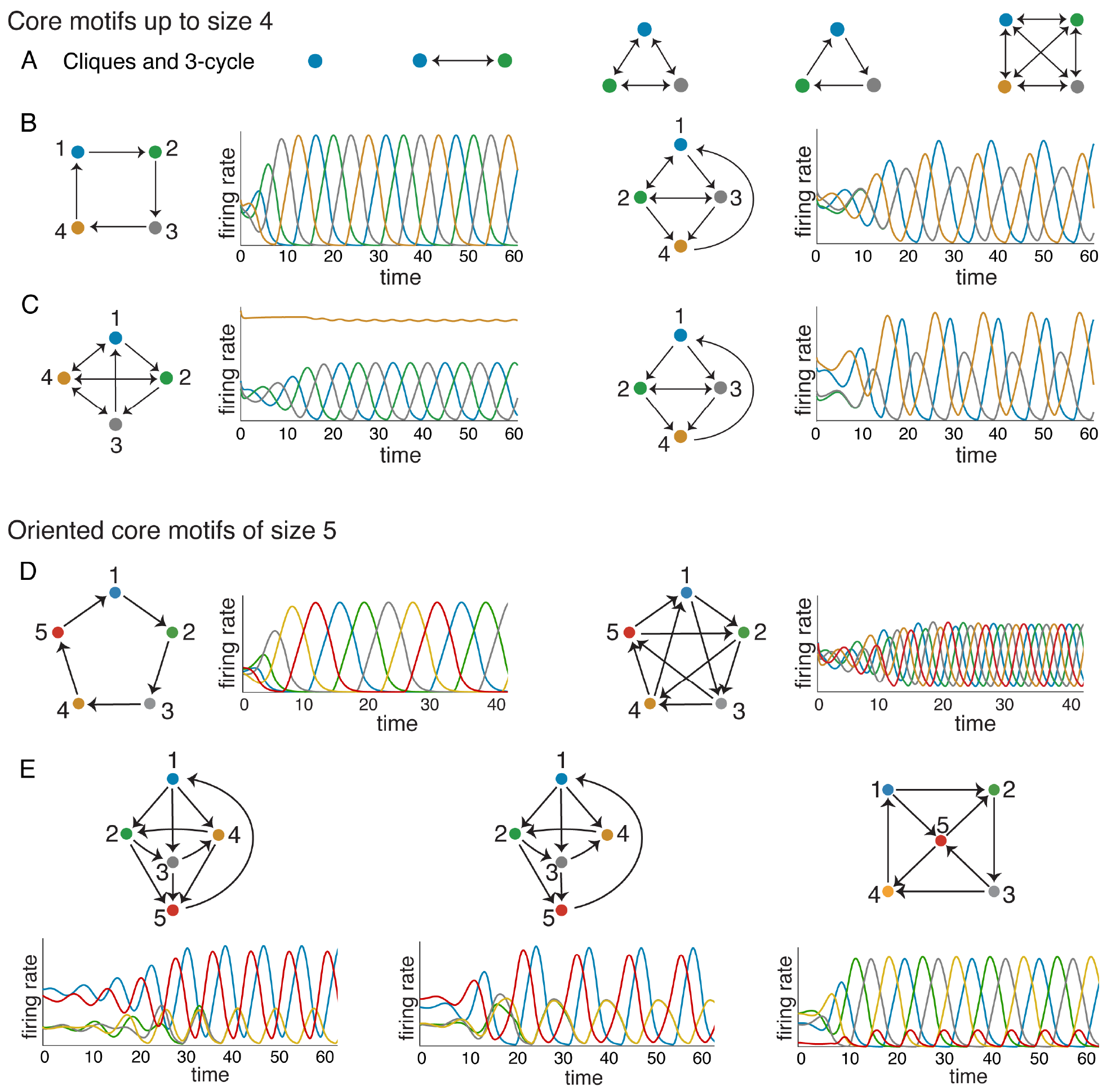}
\end{center}
\caption{{\bf Core motifs.} (A-C) All core motifs of size $n \leq 4$. Note that every clique is a core motif, as are all cycles. (B-C) Attractors are shown for each core motif of size $4$ other than the $4$-clique, whose attractor is a stable fixed point. (D-E) All $n=5$ core motifs that are {\it oriented} graphs.}
\label{fig:n4-5-core-motifs}
\end{figure}

Using graph rules, it is easy to see that all cliques (all-to-all bidirectionally connected subgraphs) and cycles of any size are core motifs, and this holds for all choices of the CTLN parameters $\varepsilon, \delta$, and $\theta$, provided they are within the legal range \cite{sequential-atts-paper}. However, there is a richer variety of core motifs beyond cliques and cycles. Fig~\ref{fig:n4-5-core-motifs}A-C depicts all core motifs up to size $n=4$. Note that all the graphs in Fig~\ref{fig:n4-5-core-motifs}A-B are uniform in-degree, but not all are cliques or cycles. The second graph in Fig~\ref{fig:n4-5-core-motifs}B has uniform in-degree 2 but no symmetry. Interestingly, its corresponding attractor exhibits a (2,3) exchange symmetry, as neurons $2$ and $3$ fire synchronously. Fig~\ref{fig:n4-5-core-motifs}C shows the two core motifs of size $n \leq 4$ that are {\it not} uniform in-degree. The first one has what we call a {\it fusion attractor}, as it appears to be a blend of the usual limit cycle supported on $123$ together with a fixed point supported on $4$. The second graph in Fig~\ref{fig:n4-5-core-motifs}C is an example of a {\it cyclic union} (see \cite{fp-paper}), with a (2,3) exchange symmetry that is reflected in its attractor. The attractors for the cliques are all stable fixed points, and the $3$-cycle attractor is the one we saw in Fig~\ref{fig:network-setup}C.

There are many more core motifs of size $n=5$. Fig~\ref{fig:n4-5-core-motifs}D-E shows all the ones that are {\it oriented}, meaning the graphs have no bidirectional edges. Fig~\ref{fig:n4-5-core-motifs}D shows the $5$-cycle (left) and the $5$-star (right), each having attractors that respect the cyclic symmetry. Fig~\ref{fig:n4-5-core-motifs}E shows the remaining three oriented core motifs and their corresponding attractors. Note that the second attractor exhibits the same (2,3,4) symmetry as the first one, even though the dropped $4 \to 5$ edge breaks the corresponding symmetry in the graph. 

Every core fixed point is supported on a core motif, but not all core motifs in a graph give rise to core fixed points. This is because the fixed point of a core motif may not survive the embedding in the larger network. For example, singletons are core motifs, but they only yield core fixed points if they are embedded as sinks in $G$. Similarly, all $3$-cycles of a graph are core motifs, but they can fail to have a surviving fixed point as in the case of $145$ in Fig~\ref{fig:rule-of-thumb}A.

The graph in Fig~\ref{fig:rule-of-thumb}C has plenty of cycles and cliques, and these are all core motifs. However, only the cliques supported on $48$ and $189$ and the $3$-cycles supported on $236$ and $345$ have surviving fixed points in $\FP(G)$. These are in fact all the core fixed points of $G$. Although $\FP(G)$ has an additional 13 fixed points (shown in Fig~\ref{fig:rule-of-thumb}C), none of them are minimal and so none can be core. By systematically trying a battery of different initial conditions in the state space, we were able to find only four attractors: two stable fixed points, a limit cycle, and a chaotic attractor. As can be seen in Fig~\ref{fig:rule-of-thumb}C (right), these attractors correspond precisely to the four core fixed points we determined using graph rules. Moreover, each core fixed point was supported on a clique or a cycle core motif. 
In other words, by identifying the core motifs of the network and applying the uniform in-degree rule, we were able to find all core fixed points via a purely graphical analysis. These core fixed points were then predictive of the network's static and dynamic attractors. 

We hypothesized that this pattern holds more generally: namely, that a network's core fixed points correspond to both its static and dynamic attractors. In the case of static attractors, our hypothesis implies that every stable fixed point is a core fixed point, and hence its support is minimal and corresponds to a core motif.  In prior work, we explored an even stronger conjecture: that every stable fixed point of a CTLN corresponds to a target-free clique \cite{stable-fp-paper}. In this work, however, we test the complementary hypothesis that dynamic attractors correspond to core fixed points that are unstable.

\subsection*{Attractor prediction}

In order to focus on dynamic attractors, we decided to study graphs whose CTLNs contain no stable fixed points, and hence no static attractors. In \cite[Theorem 2.4]{CTLN-preprint} it was shown that oriented graphs with no sinks have no stable fixed points. We thus focused our attention on oriented graphs with no sinks on $n=5$ nodes. This family is large enough to encompass a rich variety of dynamic phenomena, and small enough to be studied comprehensively. 

\subsubsection*{Oriented graphs with no sinks}
A directed graph is {\it oriented} if it has no bidirectional edges $i \leftrightarrow j$. The graph in Fig~\ref{fig:tadpole}A is oriented but has a sink, which corresponds to a static attractor. The graph in Fig~\ref{fig:rule-of-thumb}C has no sinks, but it has bidirectional edges and is thus not oriented. It has both static and dynamic attractors. In contrast, the graphs in Figs~\ref{fig:rule-of-thumb}A,B are both oriented with no sinks, as are the graphs in Fig~\ref{fig:n4-5-core-motifs}D,E. These networks are guaranteed to have only dynamic attractors, and are precisely the kind of networks we have chosen to investigate. Note that their fixed points all have support of size at least three, a useful fact that holds more generally:

\begin{lemma}\label{lemma:oriented}
Let $G$ be an oriented graph with no sinks. Then for each $\sigma \in \FP(G)$, $|\sigma| \geq 3$.
\end{lemma}

\begin{proof}
By the sinks rule, there are no singletons in $\FP(G)$. Now consider $\sigma = \{i,j\}$, with $i \to j$. By the sources rule, $\sigma \notin \FP(G)$. On the other hand, if $\sigma = \{i,j\}$ but $G\vert_\sigma$ is an independent set (no edge), then by the uniform in-degree rule for $d = 0$ we see that $\sigma \notin \FP(G)$, since $i$ and $j$ are not sinks in $G$. As there are no subsets with bidirectional edges, it follows that $\sigma \notin \FP(G)$ for all $\sigma$ of size $|\sigma| \leq 2$.
\end{proof}

\begin{figure}[!ht]
\begin{center}
\includegraphics[width=\textwidth]{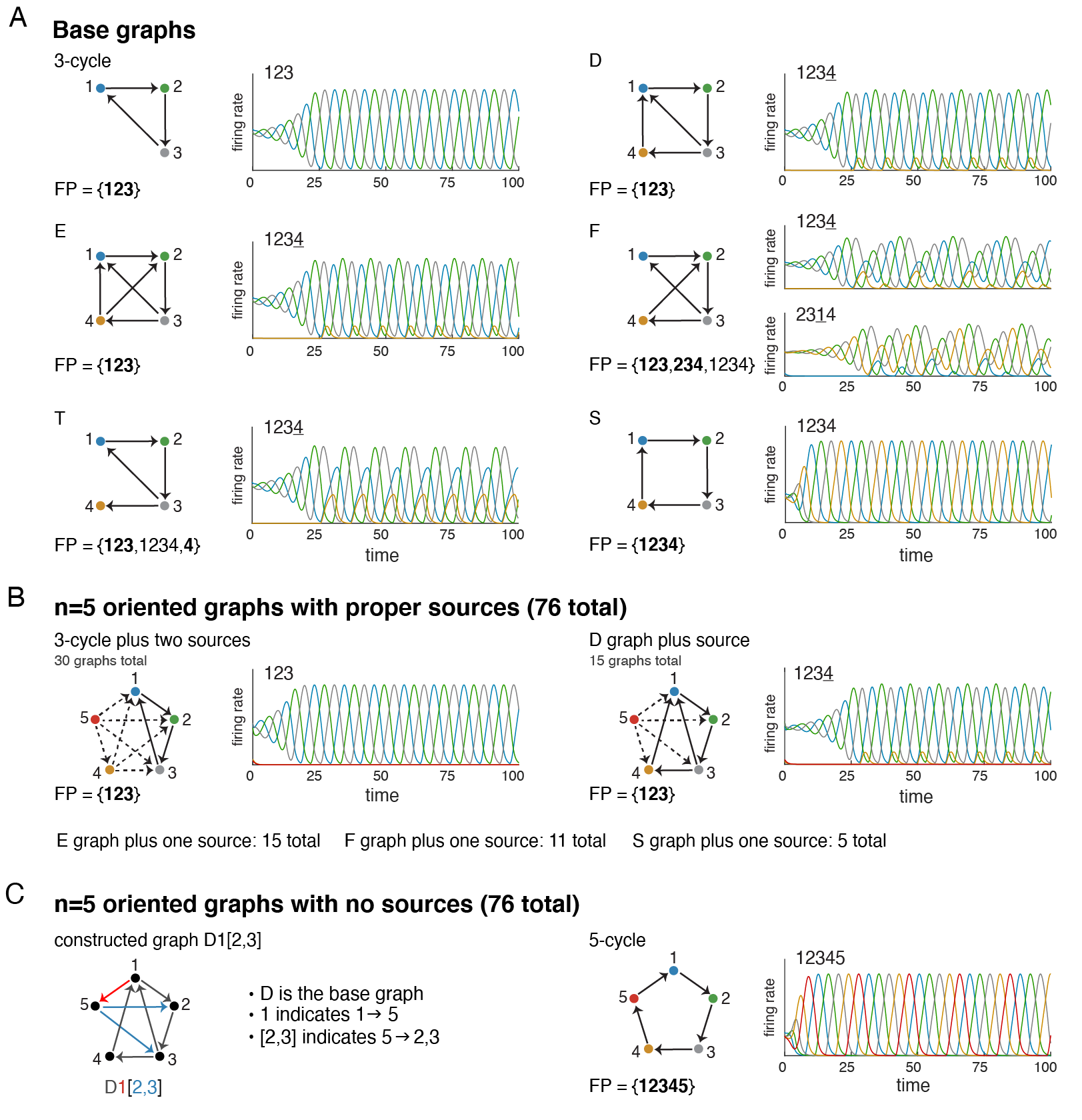}
\end{center}
\caption{{\bf Taxonomy of n=5 oriented graphs with no sinks.} (A) Base graphs used to construct $n=5$ graphs, and their corresponding attractors. Each attractor has a sequence, indicating the (periodic) order in which the neurons achieve their peak firing rates. (B) The oriented graphs with sources can be constructed by adding proper sources to each of the base graphs. This yields 30 graphs from the 3-cycle base (left), 15 graphs from the D graph base (right), and an additional 15, 11, and 5 graphs from the E, F and S graph bases. (C) All oriented graphs with no sources or sinks can be constructed from one of the D, E, F, T, and S base graphs. The graph label completely specifies the graph by naming the base and indicating the incoming and outgoing edges to the added node 5. (Left) For example, D1[2,3] is the graph constructed from the D graph with added edges $1 \to 5$ and $5 \to 2,3$. (Right) The only oriented $n=5$ graph with no sources or sinks that cannot be constructed in this way is the 5-cycle.}
\label{fig:taxonomy-setup}
\end{figure}

For $n=3$, there is only one oriented graph with no sinks: the $3$-cycle. For $n=4$, there are seven such graphs (up to isomorphism). Three of them are obtained by adding a proper source node to the $3$-cycle, with one, two, or three outgoing edges. In addition to this, there is the $4$-cycle and three more graphs obtained by adding a node to the $3$ cycle that is not a source. We call these the D, E, and F graphs (see Fig~\ref{fig:taxonomy-setup}A). In total, there are eight (non-isomorphic) oriented graphs with no sinks on $n \leq 4$ nodes.  As before, $\FP(G)$ for each of these graphs could be derived and core fixed points identified via graphical analysis. We then verified the correspondence between core fixed points and attractors computationally, for CTLNs with the standard parameters. Fig~\ref{fig:taxonomy-setup} shows the attractors corresponding to the D, E, and F graphs, as well as that of the $3$-cycle and $4$-cycle (S graph). The three graphs obtained by adding a source node to the $3$-cycle (not shown) have the same $\FP(G)$ and the same attractor as the $3$-cycle. Only the F graph has more than one core fixed point, but each one yields its own attractor, as predicted. With so few graphs, however, this was not a strong test of our hypothesis.

Fig~\ref{fig:taxonomy-setup} contains an additional T graph (a.k.a. the ``tadpole'') that is oriented, but has a sink. We include it because it is useful in our method for classifying $n=5$ oriented graphs with no sinks, described below. Only the dynamic attractor is shown, but there is also a stable fixed point supported on node $4$. Indeed, the correspondence between core fixed points and attractors for this graph was already shown in Fig~\ref{fig:tadpole}. 

Note that each attractor in Fig~\ref{fig:taxonomy-setup}A has a corresponding {\it sequence}, giving the order in which the neurons reach peak activity within a single period of the limit cycle. Underlined numbers correspond to low-firing neurons; for example, the sequence $123\underline{4}$ for the D attractor indicates that node $4$ has a low peak as compared to the other three. Since the trajectories for limit cycles are periodic, the sequence is also understood to be periodic. Our convention is to select the lowest-numbered high-firing neuron as the starting point.

\subsubsection*{Taxonomy of n=5 oriented graphs with no sinks}
For $n=5$, there are many more oriented graphs with no sinks. We developed a complete taxonomy of these graphs after splitting them into two groups: graphs with at least one source, and graphs with no sources.  There are $76$ graphs in each group (up to isomorphism), for a total of $152$ oriented graphs with no sinks. 

The graphs with sources must have {\it proper} sources, since an isolated node is also a sink. Fig~\ref{fig:taxonomy-setup}B shows the family of graphs obtained by adding node $4$ and then node $5$ as sources to the $3$-cycle. The dashed lines indicate optional edges. Keeping in mind that each added source must have out-degree at least one, we count 30 graphs in this family, up to isomorphism. The accompanying attractor is identical for every graph, and matches that of the isolated $3$-cycle in Fig~\ref{fig:taxonomy-setup}A. The second family of graphs in Fig~\ref{fig:taxonomy-setup}B comes from adding node $5$ as a proper source to the D graph. Since the D graph has no symmetry, we get $2^4-1 = 15$ distinct graphs this way. Here, too, each attractor is identical and matches that of the isolated D graph above. Fig~\ref{fig:taxonomy-setup}B shows the counts for all the other graph families obtained by adding a source. In total, there are 76 of them.

The remaining oriented graphs with no sinks have no sources. It turns out that, other than the $5$-cycle, these can all be constructed from one of the D, E, F, T, or S base graphs in Fig~\ref{fig:taxonomy-setup}A by adding node $5$ with at least one incoming and at least one outgoing edge. We developed a simple notation for these constructed graphs, which is illustrated in Fig~\ref{fig:taxonomy-setup}C. The notation uses the letter of the base graph followed by the nodes sending incoming edge(s) to node $5$, and finally the nodes receiving the outgoing edges from $5$, in brackets. For example, the graph obtained from the D graph by adding the edges $1 \to 5$ and $5 \to 2,3$ is denoted D1[2,3]. (See Supporting Information for more details.) All remaining graphs with no sources or sinks can be constructed in this manner, but many are constructible in more than one way. For example, D1[2,3] is isomorphic to E2[3]. In total, there are 75 non-isomorphic graphs obtained via this construction. Together with the $5$-cycle, these are precisely the 76 oriented graphs with no sources or sinks on $n=5$ nodes.

\subsubsection*{Parameter-independence of the attractor predictions}
From prior work \cite[Theorem 7]{fp-paper}, we know that for any oriented graph on $n \leq 5$ nodes, the set of fixed point supports $\FP(G)$ is independent of the choice of CTLN parameters $\varepsilon, \delta$, and $\theta$, provided they are in the legal range. We were thus able to completely work out $\FP(G)$ for each of the 152 graphs using graph rules, Lemma~\ref{lemma:oriented}, and a few additional facts from \cite{fp-paper} (see Supporting Information). From this graphical analysis, we also identified the core fixed points for each graph. Altogether, there were $191$ core fixed points across the $152$ graphs, with at least one core fixed point per graph. 

We hypothesized that each of these (unstable) core fixed points corresponds to a dynamic attractor, meaning that: (i) initial conditions near the fixed point yield solutions that converge to the attractor, and (ii) the support of the core fixed point predicts the high-firing neurons in the attractor. In other words, the core fixed points give us a concrete prediction for the attractors of the network, including how to find them. Note that while the exact firing rates at the core fixed points depend on parameters, their supports in $\FP(G)$ are parameter-independent. This means the prediction for the number of attractors and where they are localized within the graph of the network is the same for all parameters $\varepsilon, \delta, \theta$ in the legal range. Table~\ref{table:n5-count} provides a tally of the number of graphs and core fixed points for each subgroup of $n=5$ graphs we studied.

\subsubsection*{Testing the attractor predictions}
Next, we performed extensive searches for the attractors of the CTLNs associated to each graph, with the standard parameters. This involved simulations of network activity using a battery of initial conditions, including dozens of perturbations of each fixed point (not only core fixed points) and the 32 corners of the unit cube ${[0,1]}^5$. Remarkably, we found that every single observed attractor corresponded to a core fixed point of the CTLN, and was thus accurately predicted by graphical analysis of the network. In particular, there were no {\it spurious} attractors that were not predicted by core fixed points. On the other hand, there were six core fixed points, across six different graphs, that did not have a corresponding attractor. We refer to these predicted attractors that failed to be realized as {\it ghost} or {\it missing} attractors. The results are summarized in Table~\ref{table:n5-count}.

\begin{table}[ht]
\centering
\begin{tabular}{c|c|c|c|c|c}
 graphs & \# graphs & \# core fps & \# attractors & \# ghost atts & \# spurious atts\\
 \hline
 \hline
 with a source & 76 & 87 & 87 & 0 & 0\\
 \hline
with no source & 76 & 104 & 98 & 6 & 0\\
\hline
 total & 152 & 191 & 185 & 6 & 0\\
 \hline
\end{tabular}
\vspace{.15in}
\caption{{\bf Core fixed points and attractors for $n=5$ oriented graphs with no sinks.} The attractors were found in CTLNs with the standard parameters.}
\label{table:n5-count}
\end{table}

The attractors were predicted irrespective of the CTLN parameters, but we initially tested these predictions only in the standard parameters. Since there were ghost attractors that failed to be realized, we wondered if these attractors might emerge in a different parameter regime. For each of the six graphs with ghost attractors, we investigated CTLNs with higher inhibition levels (i.e., higher $\delta$). We found that by keeping $\theta = 1$ and $\varepsilon = 0.25$, but increasing the inhibition to $\delta = 1.25$, all six ghost attractors were observed as real attractors in the network. Moreover, all the other core fixed points continued to have corresponding attractors, and there were no new spurious attractors (see Supporting Information). In this parameter regime, the prediction of attractors from core fixed points was perfect.

\subsubsection*{Modularity of attractors}
Our taxonomy of oriented graphs with no sinks allowed us to go further in our analysis of attractors. Other than the $5$-cycle, each $n=5$ graph was constructed from a base graph of three or four nodes, where the base graph contained a core motif embedded in a canonical way. So rather than starting from a set of $152$ graphs with arbitrary orderings on the vertex labels, graphs with similarly embedded core motifs had their vertices aligned. This ensured that similarities across attractors corresponding to isomorphic core motifs were readily apparent, without having to find the optimal permutations on $x_1,\ldots,x_5$ to make the trajectories align. In particular, we expected that activation {\it sequences} associated to attractors for similarly embedded core motifs would be the same. 

Beyond such combinatorial features, however, we expected considerable variation in the precise trajectories of the attractors. After all, no two graphs are isomorphic, so the core motifs across different graphs are never embedded in exactly the same way. In particular, for a given choice of CTLN parameters, each graph yields a distinct dynamical system having a distinct $W$ matrix, with no pair of matrices being permutation-equivalent.

To our surprise, we discovered that attractors from different networks corresponding to similarly embedded core motifs were often identical, or nearly identical. Moreover, the graphs fit into simple graph families which could be described compactly via a set of common edges across all graphs in the family, together with a set of optional edges that accounted for differences between graphs. We depict these families via {\it master graphs}, with optional edges shown as dashed lines. Although the optional edges could alter $\FP(G)$, and even the number of attractors of a network, they left the aligned attractor for the graph family unchanged. Altogether we found that the 185 attractors observed in the standard parameters fell into only 25 distinct attractor classes, which we labeled att 1--25. The first attractor class, att 1, corresponds to the ``pure $3$-cycle'' attractor shown in Fig~\ref{fig:taxonomy-setup}B (left), with 30 graphs. While several attractor classes have only one graph, they vary considerably in size and the largest has 44 graphs. A full classification of attractor classes, together with master graphs, is provided in Section 4 of the Supporting Information.

\begin{figure}[!ht]
\begin{center}
\includegraphics[width=\textwidth]{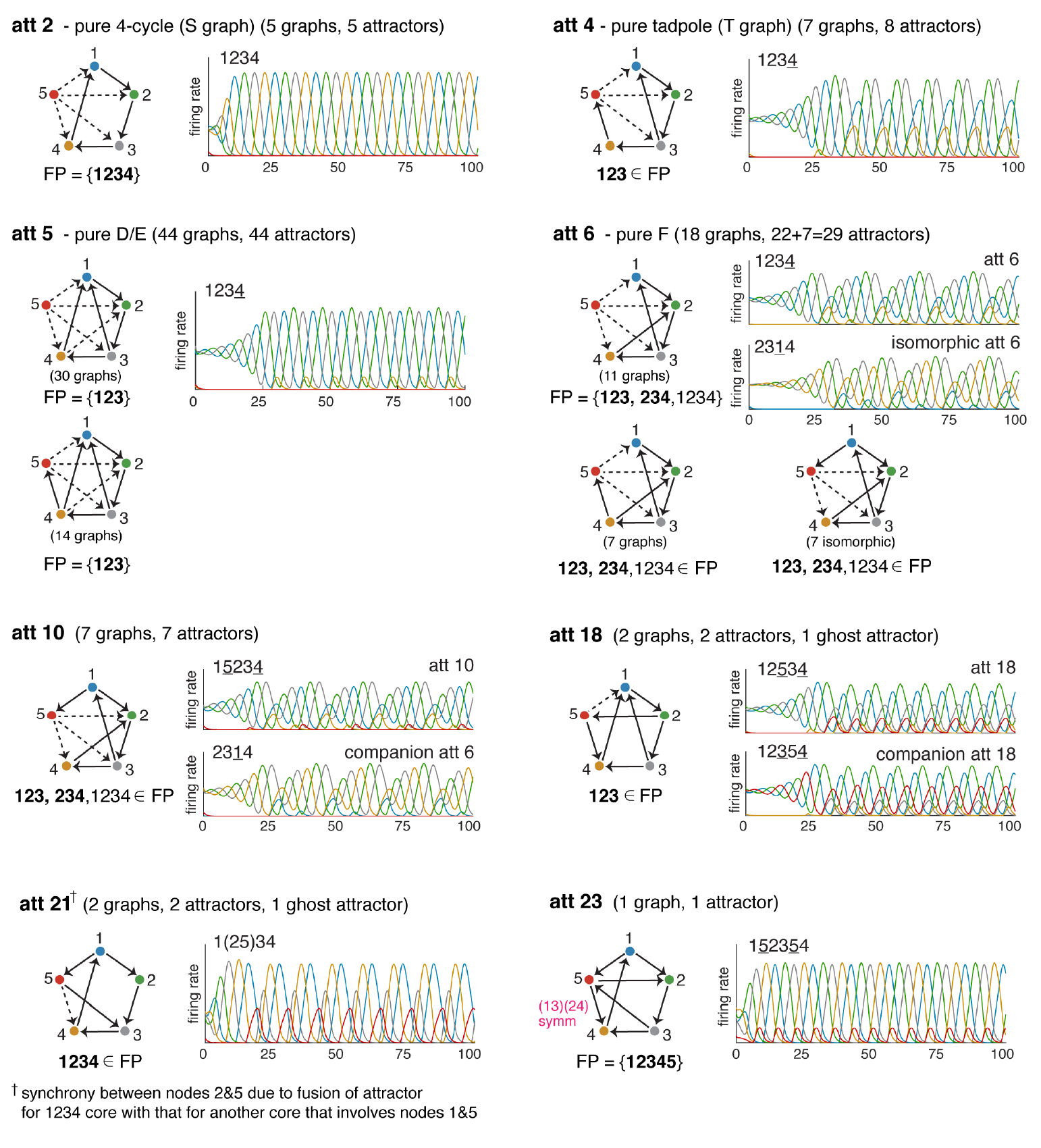}
\end{center}
\caption{{\bf Attractor classes and master graphs}. A sampling of attractor classes from the full classification for $n=5$ oriented graphs with no sinks. Each attractor emerges from multiple graphs which, once properly aligned, fit neatly into families that can be summarized by ``master graphs'' with optional edges depicted via dashed lines. For families where $\FP(G)$ is invariant across all graphs, the full form is shown. Otherwise, only the common fixed point supports are given. Some families always have two attractors: in these cases, the secondary attractor is shown as a ``companion attractor'' next to the relevant master graph. Note that the graph for att 23 has an automorphism, shown in pink. The full classification and further details of our notational conventions are provided in the Supporting Information.}
\label{fig:dictionary}
\vspace{.2in}
\end{figure}

Fig~\ref{fig:dictionary} displays eight of the attractor classes comprising 98 attractors and 2 ghost attractors. The first class shown, att 2, consists of all graphs obtained by adding a proper source to the 4-cycle. Note that since our graphs have no sinks, node $5$ must have at least one outgoing edge. Although there appear to be $2^4-1 = 15$ possibilities, many of these are permutation-equivalent and the total count is only $5$ graphs. Each of these graphs has a single attractor corresponding to the core fixed point supported on $\sigma = 1234$, as predicted. But the striking thing about these attractors is that they all appear to be identical: not only do they have the same sequence of activation, $1234$, but the rate curves look exactly the same, matching the example shown in Fig~\ref{fig:dictionary} (att 2, top left). The second class, att 4, also has optional edges coming out of node $5$. In this case, however, there is no symmetry, so all $2^3-1 = 7$ options produce non-isomorphic graphs. The reason there are $8$ attractors of this type is that one of the graphs, corresponding to choosing only the $5 \to 3$ optional edge, has a symmetry that exchanges the $123$ and $345$ cycles. This results in a second attractor, with sequence $3\underline{1}45$, that is isomorphic to the first one.

Attractor classes att 5 and att 6 have graphs that break into two families: one with $5$ as a source node, and one where $5$ receives a single edge from the base graph. Between them, they account for $44+29 = 73$ of the $191$ observed attractors. Graphs in att 5 are constructed from D and E base graphs, while the att 6 graphs all have an F graph as their base. The attractors for each family of graphs are the same, irrespective of whether the graph contains a source. For att 5, they match the isolated D and E attractors in Fig~\ref{fig:taxonomy-setup}A, and for att 6 they match the isolated F attractor. 

Note that in att 5, the bottom graph family where $5$ is not a source includes graphs where $234$, $345$, or $2345$ is a core motif (a cycle). None of these cycles can have a core fixed point, however, because node $1$ receives two edges from it. In fact, $\FP(G) = \{123\}$ for all of these graphs and they each have a single attractor. In contrast, each graph under att 6 has a secondary attractor corresponding to $234$. For the 11 graphs in the family where node $5$ is a source, this is manifested as an isomorphic att 6. The remaining 7 graphs can be represented in two ways, depending on whether we have the $4 \to 5$ edge or the $1 \to 5$ edge (by symmetry, these are equivalent). In these cases, the secondary attractor is att 10, also shown in 
Fig~\ref{fig:dictionary}. Note that although att 6 and att 10 are both supported on an embedded $3$-cycle and have a similar appearance, the attractors are fundamentally different: att 6 does not involve node $5$, while att 10 does include $5$ as a low-firing node.

The last three classes shown in Fig~\ref{fig:dictionary} are att 18, att 21, and att 23. All three of these classes contain graphs with symmetry, and this affects the attractors in different ways. In att 18, when the $5 \to 1$ edge is present the graph has a $(3, 5)$ exchange symmetry that exchanges the $123$ and $125$ core fixed points. Consequently, there are two isomorphic versions of att 18 in this graph. When $5 \not\to 1$, this symmetry is broken, and the attractor for $123$ disappears (another attractor for the $4$-cycle $1245$ emerges). We thus have a ghost attractor in the standard parameter regime. At the higher $\delta$ parameters, this attractor is realized and matches the pair for the other graph (see Supporting Information). 

\begin{table}[ht]
\centering
\begin{small}
\begin{tabular}{l|l|l|c|c|c}
attractor & sequence & graph families & \# graphs & \# attractors & \# ghosts\\
 \hline
 \hline
 att 1 & 123 & 3-cycle + sources & 30 & 30 & 0\\
  \hline
 att 2 & 1234 & 4-cycle + source (aka S0[$*$]) & 5 & 5 & 0\\
  \hline
 att 3 & 12345 & 5-cycle & 1 & 1 & 0\\
  \hline
 att 4 & 123\underline{4} & T4[$*$] & 7 & 8 & 0\\
  \hline
 att 5 & 123\underline{4} & D/E0[$*$] \& D/E4[$*$] & 30+14=44 & 30+14=44 & 0\\
  \hline
 att 6 & 123\underline{4} & F0[$*$] \& F4[$*$] & 11+7=18 & 22+7=29 & 0\\
  \hline
 att 7 & 1\underline{5}23\underline{4} & D1[2,$*$], E1[$*$], \& D/E[1,4][2,$*$] & 4+7+4=15 & 4+7+4=15 & 0\\
  \hline
 att 8 & 123(\underline{45}) &D/E3[$\sim$4,$*$] & 5 & 5 & 0\\
  \hline
 att 9 & 123\underline{54} & D/E3[4,$*$] & 8 & 8 & 0\\
  \hline
 att 10 & 1\underline{5}23\underline{4} & F1[$*$] & 7 & 7 & 0\\
  \hline
 att 11 & 12\underline{5}3\underline{4} & F2[3,$*$] \& F[2,4][3]  & 3+1=4 & 4+1=5 & 1\\
  \hline
 att 12 & 123\underline{54} & F3[1,4,$*$] & 2 & 4 & 0\\
  \hline
 att 13 & 123\underline{54} & F3[$\sim$1,4,$*$] & 2 & 0 & 2\\
  \hline
 att 14 & 123(\underline{45}) & F3[2] & 1 & 3 & 0\\
  \hline
 att 15 & 12\underline{5}3\underline{4} & F2[1,4] & 1 & 4 & 0\\
  \hline
 att 16 & 12\underline{5}3\underline{4} & E2[1,4] & 1 & 2 & 0\\
  \hline
 att 17 & 12\underline{5}3\underline{4} & E2[4] & 1 & 1 & 0\\
  \hline
 att 18 & 12\underline{5}3\underline{4} & D2[$\sim$3,4,$*$] & 2 & 2 & 1\\
  \hline
 att 19 & 1\underline{5}23\underline{4} & D1[4] & 1 & 1 & 0\\
  \hline
 att 20 & 1\underline{5}234 & S1[4] & 1 & 1 & 0\\
  \hline
 att 21 & 1(25)34 & S1[$\sim$2,3,$*$] & 2 & 2 & 1\\
  \hline
 att 22 & 1\underline{5}234 & S1[2,$*$] & 4 & 4 & 0\\
  \hline
 att 23 & 1\underline{5}23\underline{5}4 & S[1,3][2,4] & 1 & 1 & 0\\
  \hline
 att 24 & 23(\underline{154}) & E[1,3][4,$*$] & 2 & 2 & 0\\
  \hline
 att 25 & 12534 & E[1,2][3,4] & 1 & 1 & 0\\
 \hline
\end{tabular}
\end{small}
\vspace{.15in}
\caption{{\bf Graph families for attractor classes of $n=5$ oriented graphs with no sinks.} The ``$\sim$'' notation indicates a forbidden edge, while $*$s indicate optional edges. For example, D2[$\sim$3,4,$*$] represents the pair of graphs D2[1,4] and D2[4], which have no edge to node $3$. See Supporting Information for more details.}
\label{table:att-classes}
\end{table}

Instead of exchanging multiple isomorphic attractors, symmetry can also fix an attractor. This can manifest itself in two different ways: the nodes exchanged by the symmetry may have synchronous activity in the attractor, or the attractor may display a time-translation symmetry, where permuting the nodes simply shifts a trajectory in time. The latter is the kind of symmetry we see in the isolated $3$-cycle and $4$-cycle attractors. In att 21, the graph without the $5 \to 4$ edge has a $(2, 5)$ exchange symmetry, and this leads to nodes $2$ and $5$ firing synchronously in the attractor (see Fig~\ref{fig:dictionary}, bottom left). The synchronous nodes are indicated by parentheses, so that the sequence is $1(25)34.$ At higher values of $\delta$, however, this synchrony is broken and the attractor actually splits into two isomorphic limit cycles that are exchanged by the $(2, 5)$ symmetry (see Supporting Information). On the other hand, att 23 has a symmetry that manifests itself as a time shift of the trajectory, fixing the attractor without any synchrony, similar to the $3$-cycle case. 

Table~\ref{table:att-classes} gives a summary of the graph families and number of attractors for each of the 25 attractor classes. It also shows the sequence associated to each attractor. Note that different attractor classes may have the same sequence. For example, att 4, att 5, and att 6 each have the sequence 123\underline{4}. However, as can be seen in Fig~\ref{fig:dictionary}, the limit cycles are visually quite different. In the Supporting Information, we provide a complete dictionary of all the oriented graphs with no sources and sinks, together with their corresponding attractors and sequences. We also exhibit the full set of attractor classes together with diagrams of their graph families, as in Fig~\ref{fig:dictionary}.

Altogether, we have observed that in addition to core fixed points accurately predicting all the observed attractors, the embedded core motifs clustered into families of graphs that can be compactly described via a set of common and optional edges. All graphs in the same attractor class displayed identical or nearly identical attractors, even when the overall graph differed in important ways (including in the other attractors). This striking modularity of the attractors means that the same attractor, up to fine details of the dynamics, can be embedded in different networks whose dynamics may vary considerably otherwise.

\subsubsection*{Failures of attractor prediction}

\begin{figure}[!ht]
\begin{center}
\includegraphics[width=\textwidth]{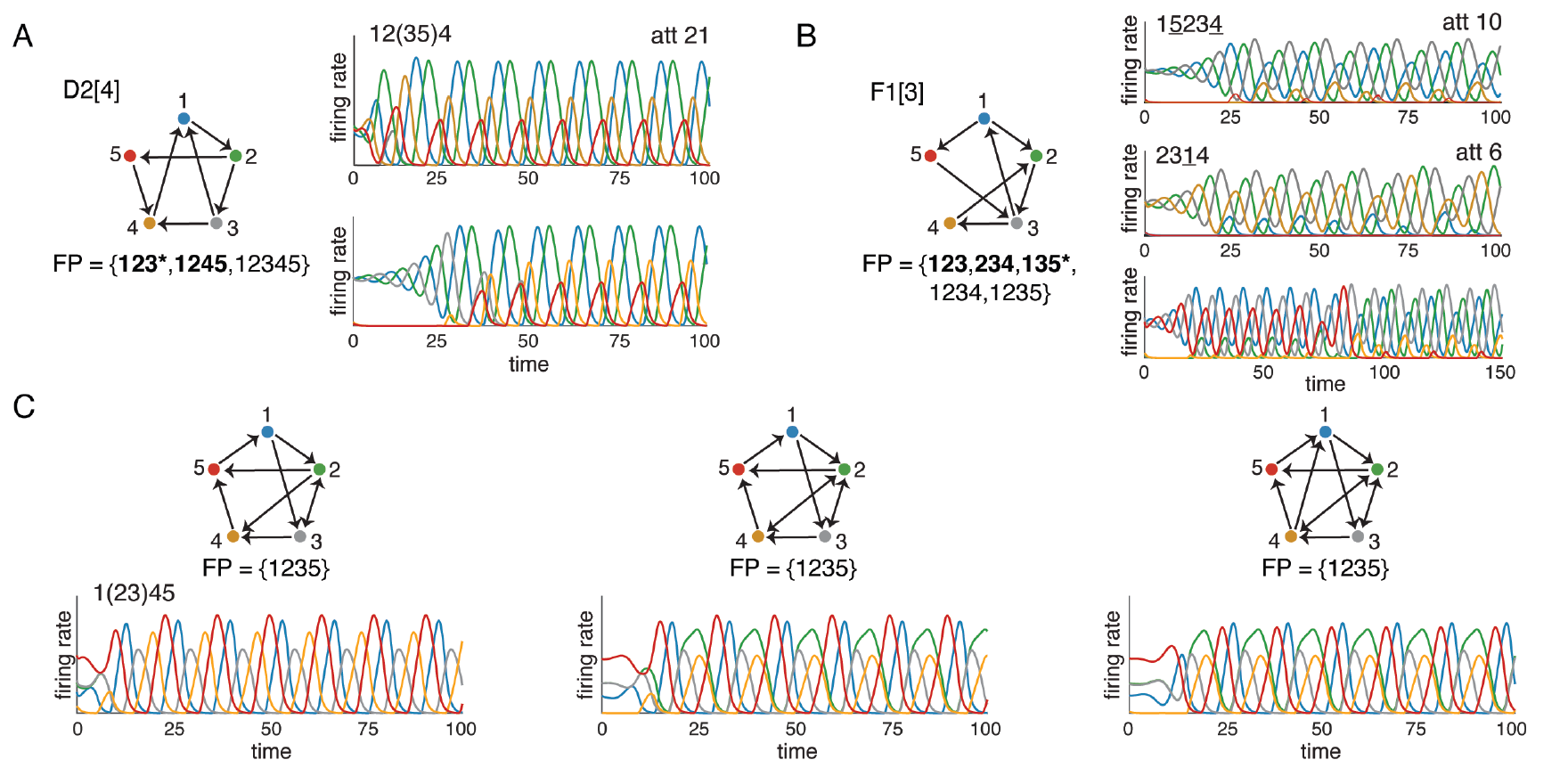}
\end{center}
\caption{{\bf Failures of attractor prediction from core fixed points.} (A) The graph D2[4] has two core fixed points, but only one attractor (att 21, top right). Initializing near the core fixed point with support $123$ leads to activity that eventually falls into the $1245$ attractor (bottom right). (B) The graph F1[3] has three core fixed points, but only the first two have corresponding attractors. Initializing near the fixed point for $135$ initially appears to fall into an attractor supported on $135$ (bottom right). However, after time these solutions converge to the attractor supported on $123$. The missing attractors in A-B are called ``ghost attractors." In a higher $\delta$ parameter regime, however, the core fixed points do yield their own attractors (see Supporting Information). (C) Three graphs that are {\it not} oriented: each one has the bidirectional edge $2 \leftrightarrow 3$. These graphs each have a unique fixed point, supported on $1235$, but it is not a core fixed point. Nevertheless, the corresponding networks all have dynamic attractors.}
\label{fig:violations}
\end{figure}

For oriented graphs with no sinks up to $n=5$ nodes, we saw that all observed attractors were predicted by core fixed points. However, there were instances of ``ghost'' attractors where a core fixed point had no corresponding attractor in the standard parameters, though these attractors were all observed in a higher $\delta$ regime. Fig~\ref{fig:violations}A illustrates what happens when initial conditions are chosen near a core fixed point with a missing attractor. The graph D2[4] has two core fixed points, supported on $123$ and $1245$. When initial conditions are chosen near the $1245$ fixed point, the solution quickly falls into the limit cycle with sequence $12(35)4$, which is isomorphic to att 21 (top right). However, when initial conditions near the $123$ fixed point are chosen, the activity initially spirals out with increasing amplitude in a $123$ sequence, but does not settle on the corresponding attractor. Instead, the activity converges to the same attractor we saw before (bottom right). At higher $\delta$, however, the analogous initial condition does produce a different attractor (see Supporting Information). Fig~\ref{fig:violations}B shows a graph where the ghost attractor is almost viable. Here it is the core fixed point for $135$ that has a missing attractor. Interestingly, initial conditions near this fixed point appear to converge to an attractor supported on $135$ (bottom right). However, the solution is not stable and eventually the activity falls out of this pattern and converges to the attractor for $123$ (top right).

Although we did have some prediction failures in the form of ghost attractors, every attractor we observed for the $n=5$ oriented graphs with no sinks was predicted by a core fixed point. There were no ``spurious'' attractors (see Table~\ref{table:n5-count}). It turns out, however, that spurious attractors can also occur as a failure of prediction. Fig~\ref{fig:violations}C shows three graphs that have no core fixed points, but still exhibit a limit cycle attractor. These are all, by definition, spurious attractors. Note that each graph contains at least one bidirectional edge, $2 \leftrightarrow 3$, and so these graphs are all outside our oriented graphs family. In each case, $\FP(G) = \{1235\}$, and this corresponds to an F graph. It is not a core motif because the F graph has three fixed points (the two minimal ones do not survive the embedding).

\begin{figure}[!ht]
\begin{center}
\includegraphics[width=\textwidth]{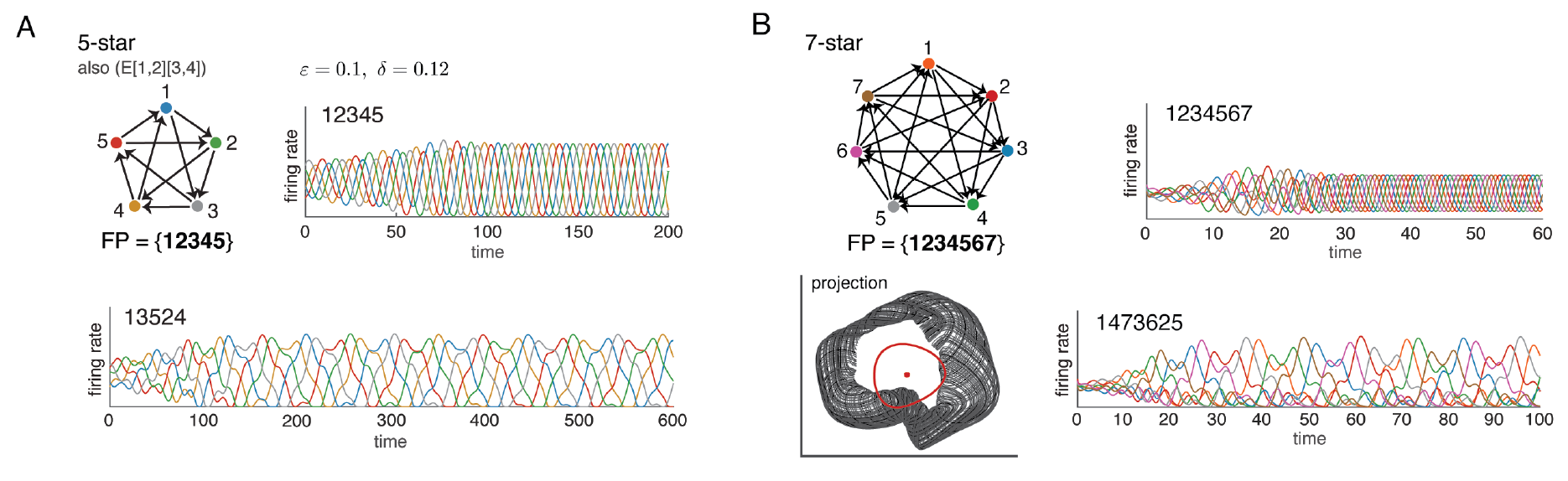}
\end{center}
\caption{{\bf Symmetry can lead to spurious attractors.} (A) Although the $5$-star graph has only a single attractor in the standard CTLN parameters, for $\varepsilon = 0.1$, $\delta = 0.12$ a second attractor emerges (bottom). Both can be accessed via small perturbations of the unique fixed point. (B) The $7$-star graph also has two attractors that can be accessed from a single core fixed point, even in the standard parameters. The projection (bottom left) depicts a random projection of $\RR^7$ onto the plane, with trajectories for the limit cycle (red circle) and an additional quasiperiodic attractor (black torus). The fixed point is also shown (red dot).}
\label{fig:5-star-7star}
\end{figure}

Finally, Fig~\ref{fig:5-star-7star} shows that symmetry can also lead to spurious attractors. This time the problem is not that there is no core fixed point, but that the same core fixed point corresponds to more than one attractor. Fig~\ref{fig:5-star-7star}A shows that this can happen even with one of our oriented $n=5$ graphs, the $5$-star. Specifically, if we go to a different parameter regime (in this case $\varepsilon = 0.1$, $\delta = 0.12$), we find that some perturbations of the core fixed point lead to the expected attractor with sequence 12345, while other initial conditions obtained by perturbing from the same fixed point lead to a very different and unusual attractor (bottom). Fig~\ref{fig:5-star-7star}B shows that a similar phenomenon occurs on the cyclically symmetric tournament on $n=7$ nodes. Here, we see two very distinct solutions corresponding to the same (and only) core fixed point. The top solution is a limit cycle, and the bottom one is quasiperiodic. The projection of the trajectories (bottom left) shows the fixed point and limit cycle in red, and the quasiperiodic trajectory with toroidal structure in black. In both networks, the graph is highly symmetric and this symmetry seems to give rise to the additional ``spurious'' attractors.

\section*{Discussion}

Predicting dynamic attractors from network structure is notoriously difficult. In this work, we have shown that in the case of CTLNs, the problem is surprisingly tractable. Specifically, we have observed a correspondence between the core fixed points of a network and its attractors. Moreover, these core fixed points have minimal supports in $\FP(G)$ and correspond to special graphs called core motifs. Using graph rules, it is straightforward to identify core fixed points in small CTLNs via structural properties of the network.

We hypothesized that core fixed points can be used to predict attractors in CTLNs. The prediction is that for each core fixed point, there is an attractor whose high-firing neurons correspond to the support of the fixed point, and the attractor can be accessed by initial conditions that are small perturbations of the fixed point. In the case of stable fixed points, which arise for core motifs that are cliques, this prediction trivially holds. So we set out to test the hypothesis for unstable core fixed points, which give rise to dynamic attractors. We focused on oriented graphs with no sinks, as these networks are guaranteed to have only unstable fixed points. 

Out of 152 oriented graphs with no sinks on $n=5$ nodes, we observed 185 attractors. All of them were predicted by core fixed points. Moreover, we found that the attractors clustered into only 25 attractor classes, with attractors in the same class being nearly identical. We were also able to organize the graphs having the same attractor into simple structural graph families. These graph families highlight the close connection between structural properties of embedded core motifs and the resulting dynamic attractors. In particular, the same attractor can be embedded in different networks whose dynamics are completely different outside the common attractor.

Our attractor prediction was not perfect, however. In the standard parameter regime, we also had 6 failures in the form of ghost attractors, which were predicted but not realized by the network. We also saw examples of networks with no core fixed points, that nevertheless had dynamic attractors. Finally, we observed that highly symmetric networks can have core fixed points that give rise to multiple attractors. Despite these caveats, we conclude that core motifs and core fixed points are important tools for connecting network structure to dynamics. And in small networks, the attractor predictions from core fixed points are surprisingly accurate.

\section*{Acknowledgments} This work was supported by NIH R01 EB022862, NIH R01 NS120581, NSF DMS-1951165, and NSF DMS-1951599.  We thank Carolyn Shaw for earlier contributions to the analysis of fixed points of CTLNs, which helped motivate the concept of core fixed points.

\newpage


\part*{Supporting Information} 
\noindent {\Large{\bf Taxonomy of attractors for oriented graphs on n=5 nodes}}\\

In this supplement we describe a classification scheme for oriented graphs with no sinks on $n=5$ nodes and their attractors. Recall that a directed graph is {\it oriented} if it has no bidirectional edges, and it has {\it no sinks} if each node has at least one outgoing edge. We have chosen to study these graphs because their corresponding CTLNs are guaranteed to not have any stable fixed points \cite[Theorem 2.4]{CTLN-preprint}. This enables us to focus our attention on testing the correspondence between core fixed points and dynamic attractors. 

Our labeling method for oriented graphs relies on {\it base graphs} of smaller size. It aligns the vertices of core motifs that are embedded in similar ways. In particular, attractors stemming from the same core motif are aligned across graphs, so that their similarities and differences are more salient. Using this labeling scheme, we provide a dictionary of oriented $n=5$ graphs, together with their corresponding attractors in the CTLN with standard parameters, $\varepsilon = 0.25$ and $\delta = 0.5$. This dictionary allowed us to classify the attractors based on sequence structure and visual similarity. The classification identified 25 distinct attractor classes from the 185 observed attractors. Moreover, we found that graphs supporting the same attractor clustered into highly structured graph families.

\tableofcontents

\section{Base graphs and graph counts}
The smallest oriented graph with no sinks is a $3$-cycle. This implies that every oriented graph with no sinks on $n \leq 5$ nodes must be connected, since each connected component must have at least $3$ nodes. We can split the $n = 5$ oriented graphs with no sinks into two parts: graphs with sources and graphs without any sources.

\begin{figure}[!ht]
\begin{center}
\includegraphics[width=\textwidth]{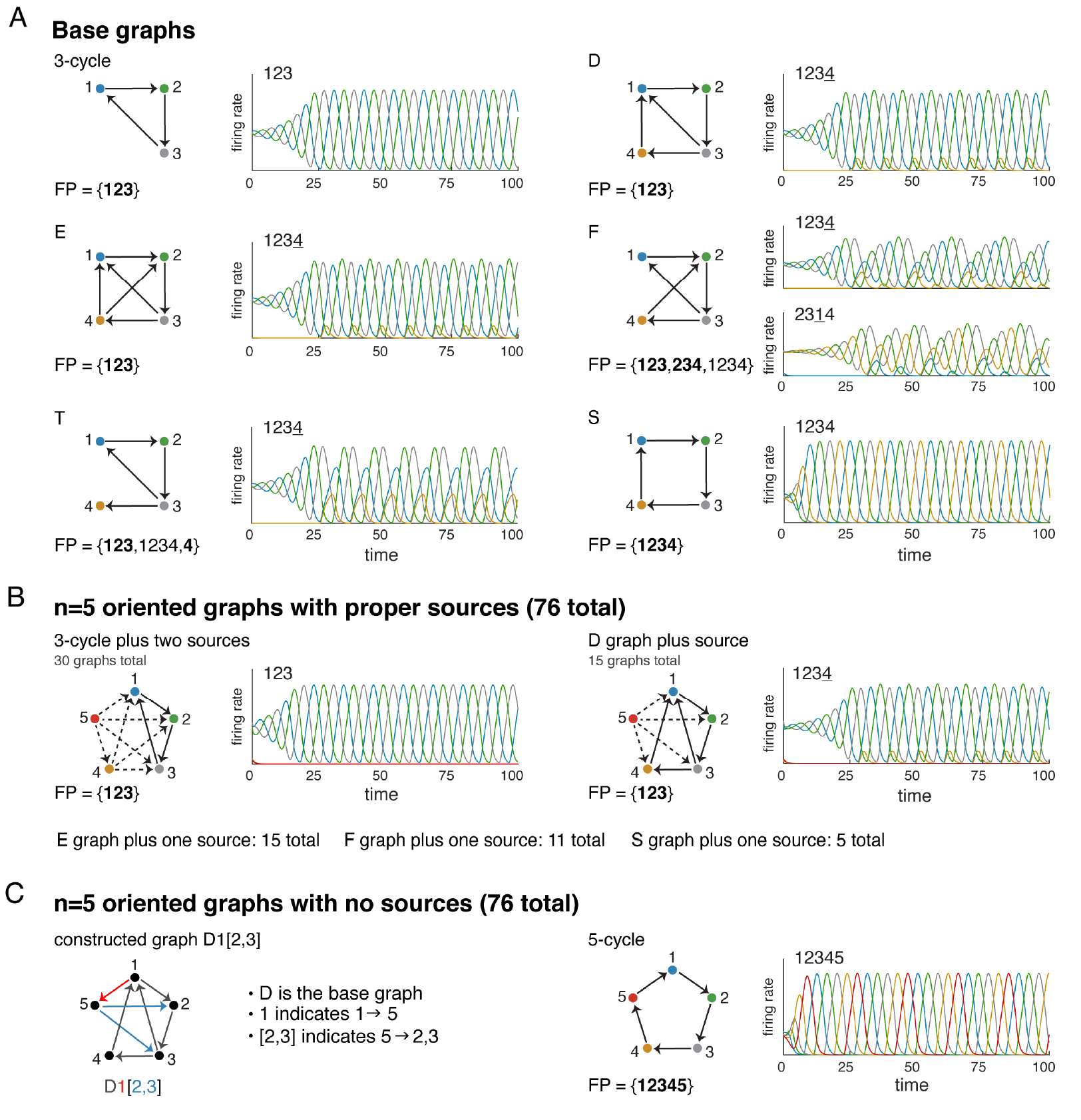}
\caption{{\bf Base graphs and graph counts.} (A) Base graphs used to construct $n=5$ graphs, and their corresponding attractors. Each attractor has a sequence, indicating the (periodic) order in which the neurons achieve their peak firing rates. (B) The oriented graphs with sources can be constructed by adding proper sources to each of the base graphs. This yields 30 graphs from the 3-cycle base (left), 15 graphs from the D graph base (right), and an additional 15, 11, and 5 graphs from the E, F and S graph bases. (C) All oriented graphs with no sources or sinks can be constructed from one of the D, E, F, T, and S base graphs. (Left) For example, D1[2,3] is the graph constructed from the D graph with added edges $1 \to 5$ and $5 \to 2,3$. (Right) The only oriented $n=5$ graph with no sources or sinks that cannot be constructed in this way is the 5-cycle. (Same as Fig 6 in the main text.)}
\label{fig:taxonomy-setup-supp}
\end{center}
\vspace{-.2in}
\end{figure}

There are a total of $152$ oriented graphs with no sinks on $n=5$ nodes.
We have verified that, with the exception of the $5$-cycle, all of these graphs can be constructed by adding a single vertex to one of the five base graphs D, E, F, T, S, shown in Fig~\ref{fig:taxonomy-setup-supp}A, or two vertices to the $3$-cycle.  Since the D, E, F, and T bases cover all cases where there is an outgoing edge from the $3$-cycle, the $3$-cycle base is only needed for graphs where node $5$ is a source, and node $4$ is a source upon removal of $5$. Note that the ``tadpole'' graph T has a sink node, $4$, so any construction using this base must have a $4 \to 5$ edge.

In total, there are 76 graphs obtained by adding node $5$ as a proper source\footnote{Recall that a {\it proper} source is a vertex with no incoming edges and at least one outgoing edge.} to one of the base graphs, 30 of them using the $3$-cycle base (see Fig~\ref{fig:taxonomy-setup-supp}B).  Note that we are counting non-isomorphic graphs, and symmetries make many of the counts nontrivial. When a base graph has no symmetry, as with the D and E graphs, there are $2^4-1 = 15$ distinct ways to add node $5$ as a  proper source (with at least one outgoing edge). On the other hand, when the base graph has symmetry, as in the F and S graphs, some of the choices for outgoing edges from the added node $5$ are isomorphic, and the count is thus lower.

An additional 75 graphs constructed from a base graph do not have sources. Together with the $5$-cycle, there are 76 oriented graphs with no sources or sinks (see Fig~\ref{fig:taxonomy-setup-supp}C). These are the graphs we focus on for the labeling scheme and dictionary. Graphs with sources are simpler to classify: their attractors are identical to those of the corresponding base graphs in Fig~\ref{fig:taxonomy-setup-supp}A.

\FloatBarrier

\section{Constructed graphs and the labeling scheme}

\begin{figure}[!ht]
\vspace{-.1in}
\begin{center}
\includegraphics[width=.8\textwidth]{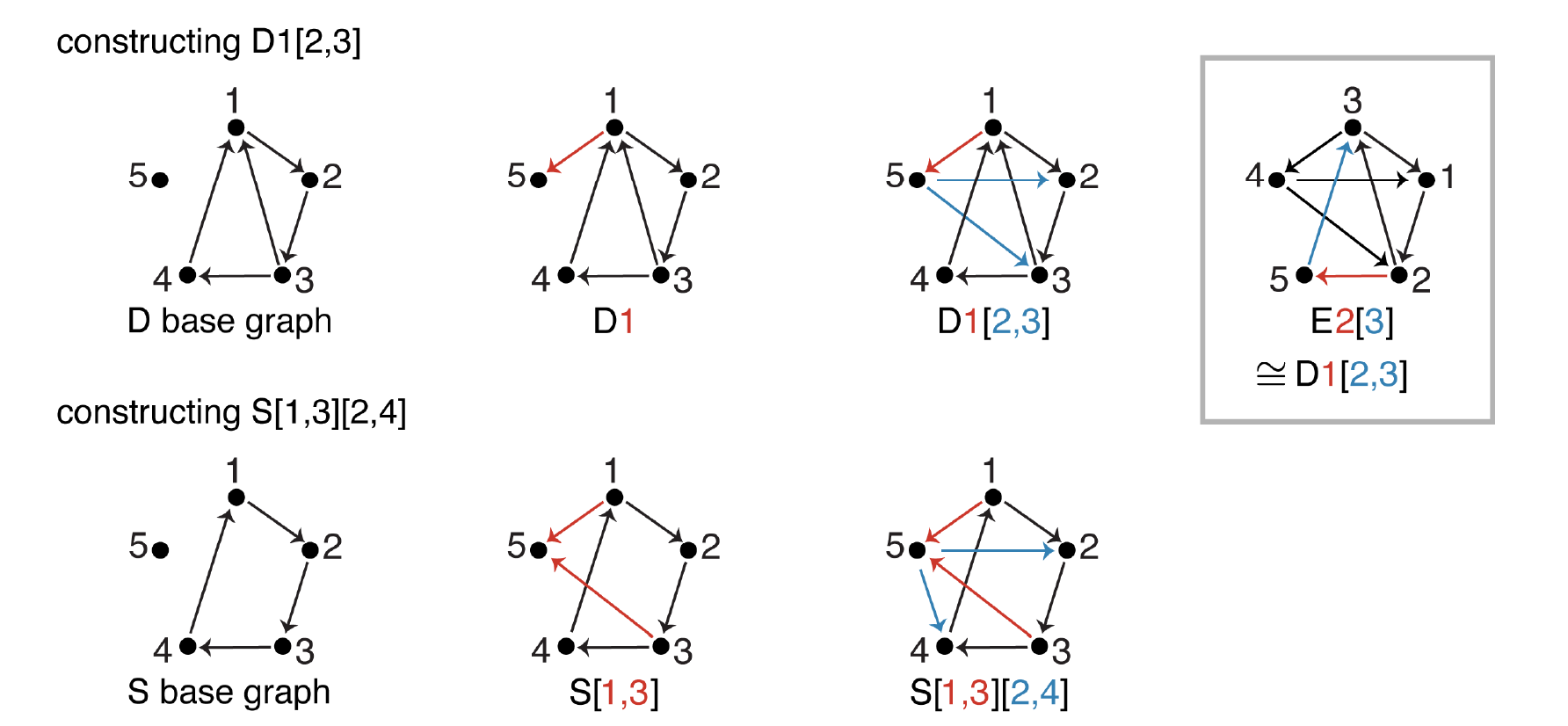}
\vspace{.1in}
\caption{{\bf Construction of oriented graphs from base graphs.} (Top) Starting with a D base, the graph D1[2,3] is constructed by adding a node $5$ together with incoming edge $1 \to 5$ (red) and outgoing edges $5 \to 2$ and $5 \to 3$ (blue). An isomorphic graph, E2[3], can be constructed from an E base. (Bottom) The graph S[1,3][2,4] has two incoming edges to node $5$, given in the first set of brackets. This graph cannot be constructed from any base with only one edge into node $5$.}
\label{fig:labeling-scheme}
\end{center}
\vspace{-.15in}
\end{figure}

\begin{figure}[!ht]
\vspace{-.1in}
\begin{center}
\includegraphics[width=.6\textwidth]{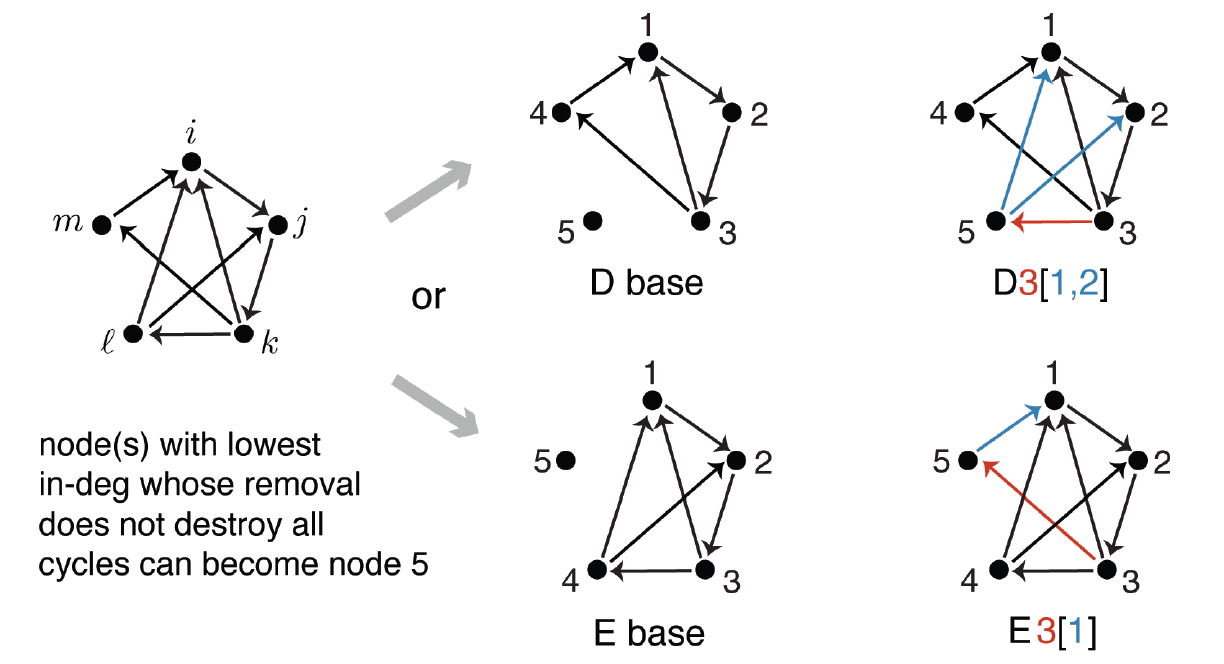}
\caption{{\bf Finding the name for a given oriented graph with no sinks.} The nodes with lowest in-degree are $k$, $\ell$, and $m$.  However, removing $k$ results in a graph with no cycles that cannot match one of our base graphs. Removing $\ell$ (top) uncovers a D graph base, while removing $m$ (bottom) results in an E base. The original graph can thus be labeled as D3[1,2] or E3[1].}
\label{fig:naming-graph}
\end{center}
\vspace{-.25in}
\end{figure}

We have devised a labeling scheme for all the $n=5$ graphs that can be constructed by adding a fifth node to one of the base graphs D, E, F, T, or S. 
We label each graph compactly with a letter and two sets of numbers indicating which nodes connect to $5$ (see Fig~\ref{fig:taxonomy-setup-supp}C). The letter indicates the base graph, which is the induced subgraph on nodes $1$ to $4$.  The number(s) immediately following the letter indicate nodes that send edges to $5$; and the second set of numbers, in brackets, indicate nodes that receive edges from $5$. Since these graphs are all oriented, a node cannot both send and receive an edge to $5$, so the two sets of numbers are disjoint. They are also nonempty, since node $5$ is neither a sink nor a source. Fig~\ref{fig:labeling-scheme} illustrates the labeling scheme for the graphs D1[2,3] and S[1,3][2,4]. Fig~\ref{fig:naming-graph} depicts how, given a graph, one can find its label(s).

Notice that constructions with different bases can produce isomorphic graphs.  For example, we see that D1[2,3] $\cong$ E2[3] in Fig~\ref{fig:labeling-scheme} and D3[1,2] $\cong$ E3[1] in Fig~\ref{fig:naming-graph}, under appropriate relabelings of the nodes. In order to maximally align our attractors and take advantage of the common base graph structures, we prioritize representations that minimize the number of edges to node $5$. 
It turns out that a labeling where node $5$ has in-degree 0 or 1 exists for all but four graphs: E[1,2][3,4], E[1,3][2,4], E[1,3][4], and S[1,3][2,4].\footnote{In these four graphs, the core motif of the base graph does not survive the addition of node $5$, thus producing a new core motif of size $n=5$.} We can also eliminate some redundancy for constructions with an F graph. Since F has a (1,4) exchange symmetry, it follows that graphs of the form F1[$*$] are isomorphic to graphs of the form F4[$*$]. Similarly, F2[1] is isomorphic to F2[4]. In our dictionary of graphs, we choose $1$ over $4$ whenever possible.

With these restrictions, our construction method produces 91 oriented graphs with no sinks and no sources (excluding the 5-cycle). We have not eliminated all redundancy, however, and 16 of these graphs appear twice under two different names. Our construction thus accurately recovers all 75 non-isomorphic oriented graphs with no sinks and no sources, other than the 5-cycle.  

Table~\ref{table:dictionary} displays families of constructed graphs, with and without sources, organized by base graph. Every oriented $n=5$ graph with no sinks is covered by one (or more) of these families. The notation for graph families works as follows. The families with a ``0" in the label, D0[$*$], E0[$*$], etc., indicate that a source was added to the base graph. The [$*$] indicates that edges out of node $5$ are all optional, and can occur in any possible combination provided at least one is selected. Note that T0[$*$] consists of graphs with node 4 as a sink, which do not belong to our oriented graphs with no sinks family. We include it because it is the first graph family where att 4 emerges, but it does not count towards the total number of graphs with sources.

\begin{table}[!ht]
\centering
\begin{tabular}{l|c|l}
graph family & \# graphs & attractor classes\\
\hline
\hline
3-cycle + sources & 30 & att 1\\
\hline
4-cycle + source (aka S0[$*$]) & 5 & att 2\\
\hline
5-cycle (no source) -- core motif & 1 & att 3\\
\hline
T0[$*$] (note: all have sinks!) & 15 & att 4\\
\hline
D0[$*$] & 15 & att 5\\
\hline
E0[$*$] & 15 & att 5\\
\hline
F0[$*$] & 11 & att 6\\
\hline
D1[$*$] & 7 & att 4, 7, 11, 17, 19\\
\hline
D2[$*$] & 7 & att 6, 7, 10, 18, 21\\
\hline
D3[$*$] & 7 & att 8, 9\\
\hline
D4[$*$] & 7 & att 5\\
\hline
E1[$*$] & 7 & att 7\\
\hline
E2[$*$] & 7 & att 6, 7, 10, 11, 16, 17\\
\hline
E3[$*$] & 7 & att 8, 9\\
\hline
E4[$*$] & 7 & att 5\\
\hline
F1[$*$] & 7 & att 6, 10\\
\hline
F2[$*$] & 5 & att 6, 7, 10, 11, 15\\
\hline
F3[$*$] & 5 & att 8, 12, 14\\
\hline
T4[$*$] & 7 & att 4, 19, 20\\
\hline
S1[$*$] & 7 & att 4, 20, 21, 22\\
\hline
S[1,3][2,4] -- core motif & 1 & att 23\\
\hline
E[1,3][4,$*$] -- core motifs & 2 & att 24\\
\hline
E[1,2][3,4] -- core motif & 1 & att 25\\
\hline
\hline
total with sources (excluding T0[$*$]) & 76 & att 1, 2, 5, 6\\
\hline
total with no sources (includes duplicates) & 92 & atts 3--25\\
\hline
total with no sources -- non-isomorphic & 76 & atts 3--25\\
\hline
\end{tabular}
\vspace{.15in}
\caption{Graph families organized by base graph.}
\label{table:dictionary}
\end{table}

The remaining families have labels such as D2[$*$] and E[1,3][4,$*$], which consists of graphs that do not have sources. The graph family D2[$*$] consists of graphs with the $2 \to 5$ edge, with the outgoing edges $5 \to 1, 3, 4$ all optional. It is understood that $5 \not\to 2$ because there can be no bidirectional edges, while at least one outgoing edge must be selected since $5$ is not a sink. In the case of E[1,3][4,$*$], the $*$ indicates that $5 \to 2$ is optional, as that is the only possibility left.

For each graph family in Table~\ref{table:dictionary}, the number of graphs is easy to count when one takes symmetry into account and recalls that node $5$ must have at least one outgoing edge. Note also that there is no F4[$*$] family listed, since it is isomorphic to F1[$*$]. Similarly, there are no S2[$*$], S3[$*$], or S4[$*$] families, since they are all isomorphic to S1[$*$]. Finally, T4[$*$] is the only T graph family without sinks, since we must have the $4 \to 5$ edge whenever T is the base graph.

The attractor classes in Table~\ref{table:dictionary} were defined by first enumerating the distinct attractors arising in the base graphs, together with the $5$-cycle. Note that the D and E graphs have the same attractor (see Fig~\ref{fig:taxonomy-setup-supp}A). The remaining attractor classes were defined from the dictionary, given in the next section, by introducing a new number for each new type of attractor that arose.

\section{Dictionary of n=5 oriented graphs with no sources or sinks}

It follows from the {\it sources graph rule} (see Table 1 of the main text) that proper sources cannot participate in fixed point supports. Moreover, it was proven in \cite{fp-paper} that if $i$ is a proper source in $G$, then $\FP(G) = \FP(G|_{[n]\setminus i})$. In other words, proper sources can be iteratively removed from a graph without altering the fixed points of the corresponding CTLN. This means that for graphs obtained by adding a source node to one of the base graphs, the set of fixed point supports $\FP(G)$ exactly matches that of the base. As it turns out, the attractors in the S0[$*$], T0[$*$], D0[$*$], E0[$*$], F0[$*$], and $3$-cycle + sources families also exactly match the attractors of the corresponding base graphs, which are shown in Fig~\ref{fig:taxonomy-setup-supp}A. Together with the $5$-cycle, this covers the first seven lines of Table~\ref{table:dictionary}.

The remaining 92 graphs have no sources or sinks, and all but the $5$-cycle can be constructed from one of the base graphs D, E, F, T, and S. We have already seen the attractor for the $5$-cycle in Fig~\ref{fig:taxonomy-setup-supp}C. In the remainder of this section, we provide a complete dictionary for the 91 graphs in Table~\ref{table:dictionary}, starting with graph family D1[$*$] (line 8) and ending with E[1,2][3,4] (line 23). Some of these are isomorphic copies of the same graph, with different base graphs corresponding to different permutations of the nodes. It is useful to view these in both representations, however, as they can ``center'' the graph on different core fixed points. In particular, most of the graphs with multiple attractors have more than one base graph representation. Note, however, that in the case of the F graphs we do not include isomorphic copies that stem from the obvious (1,4) exchange symmetry. In particular, we do not include F4[$*$] graphs in our dictionary, just as they were not included in Table~\ref{table:dictionary}. We also remove isomorphic copies from the F2[$*$] and F3[$*$] families that are due to the (1,4) symmetry, which is why these families have only 5 graphs each, as opposed to 7.

Our dictionary consists of eight pages of graphs, two for each of the D, E, and F graph families, one for the S and T families (combined), and one for the constructed core motifs: S[1,3][2,4], E[1,3][4], E[1,3][2,4], and E[1,2][3,4], which appear in the bottom three rows of Table~\ref{table:dictionary}. (Note that the core motifs must all have two incoming edges to node $5$, as the added node must ``kill'' the core fixed point for the cycle in the base graph.) 

From \cite[Theorem 7]{fp-paper}, we know that $\FP(G)$ for any oriented graph on $n \leq 5$ nodes is parameter-independent, provided $\varepsilon$ and $\delta$ are in the legal range. For each graph, we were thus able to compute $\FP(G)$ using graph rules. Furthermore, we identified the core fixed points, which all have minimal support. Their supports are indicated in bold. 

We then simulated solutions to the corresponding CTLN with standard parameters, $\varepsilon = 0.25$ and $\delta = 0.5,$  and systematically searched for attractors using a battery of dozens of initial conditions, the same for all graphs. To this we added initial conditions that were perturbations from each fixed point (not only core fixed points). We plotted the resulting solutions up to time 100 in units of the leak timescale, $\tau$, which has been normalized to $1$ in our TLN equations. All solutions converged quickly to an attractor in this time. Moreover, each observed attractor could be accessed via an initial condition that was a perturbation of a core fixed point. In the following dictionary pages, we selected example solutions obtained from this set, so that the transient activity shows how the activity spirals away from the core fixed point before converging to the attractor. All observed attractors are displayed next to each graph. In the case of F3[2] and F2[1,4], there are three and four isomorphic attractors, respectively. Here we showed only two examples of each, as the rest could be inferred by symmetry. 

We classified attractors systematically as follows: for each new graph, we examined all attractors and compared them to the previously observed attractor classes. This often entailed permuting the variables, $x_1(t), \ldots, x_5(t),$ until two attractors were maximally aligned. We examined firing rate curves, as shown in the following dictionary pages, and also random projections of the trajectories in $\RR^5$. Only when agreement between two attractors was near perfect did we decide to cluster them into the same class. Altogether, we observed 97 attractors across the 75 non-isomorphic oriented graphs listed in our dictionary pages. While some of them were repeats of attractor classes 4-6, we also identified 19 new classes. After reordering these classes to give nearby numbers to similar attractors, we came up with final names for them as att 7-25. These attractor labels are shown in the upper right corner of each firing rate plot.

Note that secondary attractors superficially look different (e.g., the colors on the rate curves don't match) because they are centered on a different permutation, where the core fixed point is not supported on
$123$ or $1234$. Moreover, even when the core fixed points have the same support, attractors may not be fully aligned. For example, D1[2] and D2[3] are isomorphic and each have a single core fixed point supported on $123$. The attractors have different sequences, though, which may lead one to believe they are distinct. However, there is a permutation that takes the labeled graph for D2[3] to D1[2], and this permutation reveals that the attractors are in fact identical and have the same sequence (as expected, since the graphs are isomorphic). Similarly, there is a permutation that realigns the attractor for D2[1,3], as it has the same sequence as D2[3]. All three of these graphs have the same attractor, att 7.

Altogether, across the 75 non-isomorphic graphs in our dictionary, there were 103 core fixed points. We thus predicted 103 dynamic attractors. However, we only observed 97 attractors.\footnote{When comparing to Table 2 in the main text, recall that the ``with no sources'' line in the table includes the $5$-cycle, so it has one extra graph, core fixed point, and attractor.} Although every observed attractor was predicted by a core fixed point, there were 6 instances where a core fixed point did not have a corresponding attractor. We refer to these as {\it ghost attractors}, because they failed to be realized in the standard parameters. Each graph that has a ghost attractor is labeled with a $**$ in the dictionary, and the core fixed point with the missing attractor has a superscript $^*$. For example, on the first page, the graph D2[4] has a ghost attractor corresponding to the core fixed point with support $123$. At the end of this section, we return to these ghost attractors and show that they can be realized by CTLNs in a higher $\delta$ parameter regime.

The vast majority of graphs have only one or two core fixed points, with at most two dynamic attractors that are limit cycles. Only three graphs have more than two core fixed points. They are each somewhat exotic:
\begin{itemize}
\item F3[2] has 3 core fixed points and 3 quasiperiodic attractors, all isomorphic to each other.
\item F2[1,4] has 4 core fixed points and 4 chaotic attractors, all isomorphic to each other.
\item F1[3] has 3 core fixed points, but one leads to a ghost attractor. This graph has three $3$-cycles, but they are not symmetrically embedded. The $123$ and $234$ cycles yield distinct attractors, while the $135$ cycle corresponds to a ghost attractor.
\end{itemize}
Finally, for graphs that have nontrivial automorphism groups we indicate the symmetries in pink.

\newpage
\subsection*{D graphs: D1[$*$] \& D2[$*$]}
\begin{center}
\includegraphics[width=\textwidth]{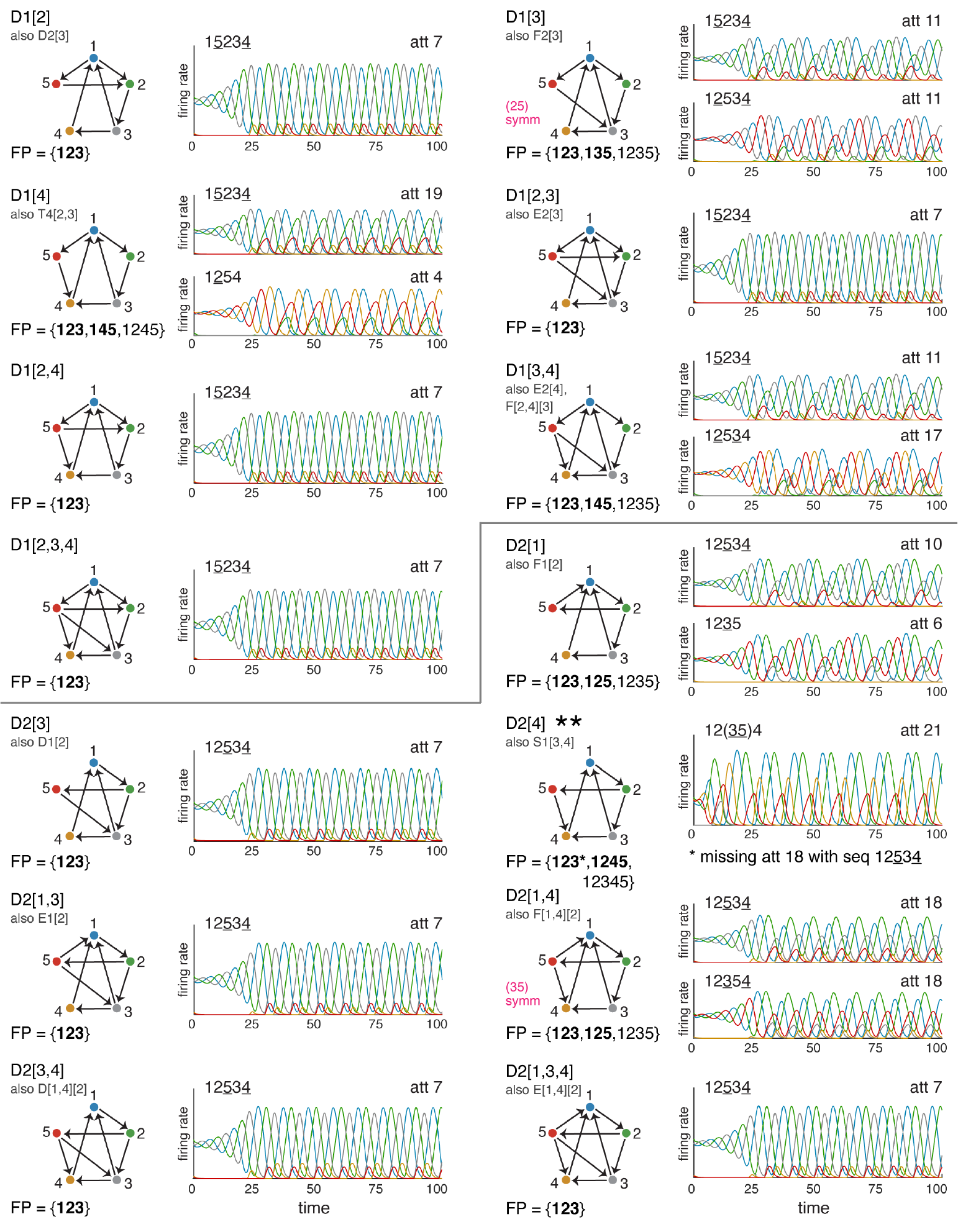}
\end{center}

\newpage
\subsection*{D graphs: D3[$*$] \& D4[$*$]}
\begin{center}
\includegraphics[width=\textwidth]{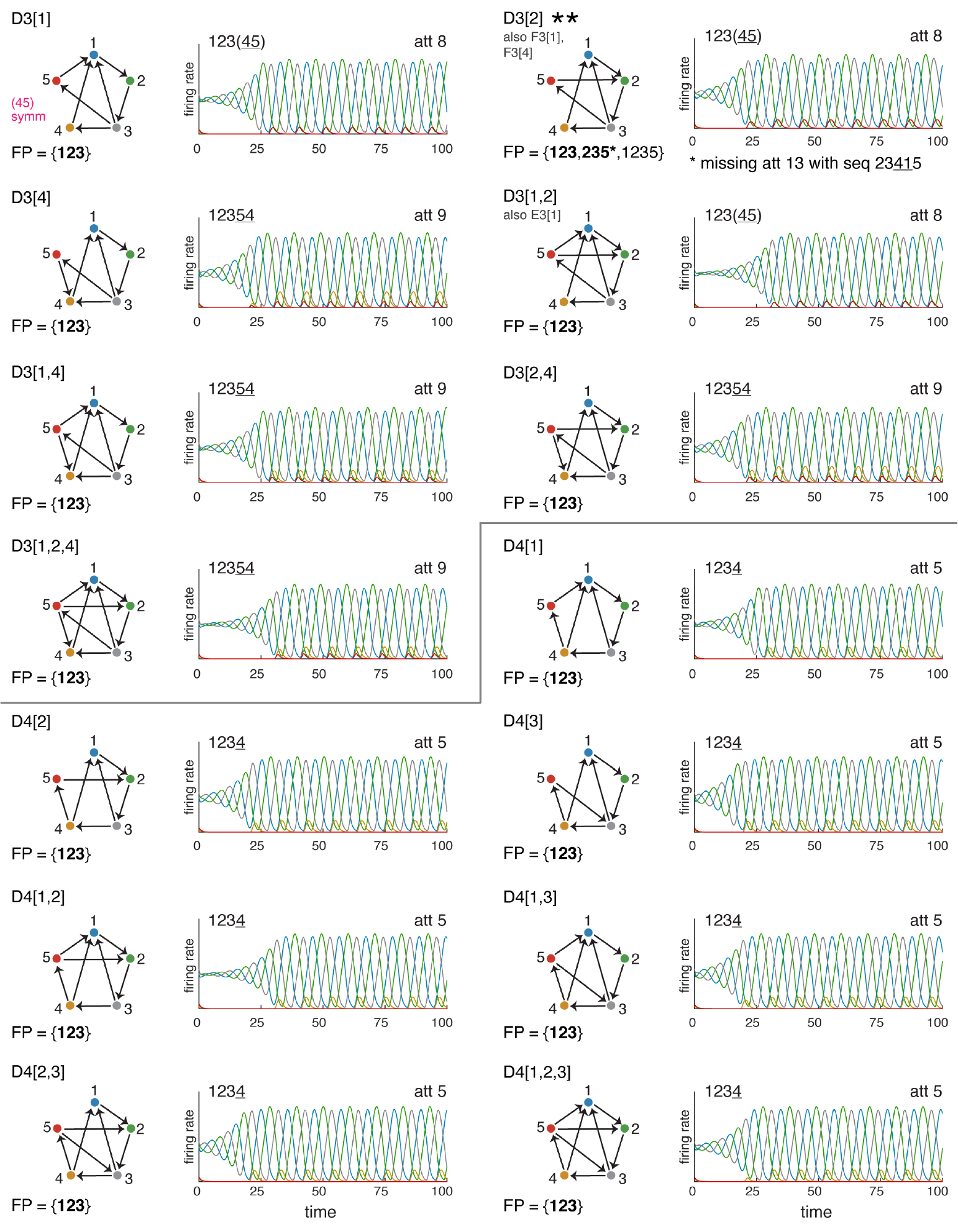}
\end{center}

\newpage
\subsection*{E graphs: E1[$*$] \& E2[$*$]}
\begin{center}
\includegraphics[width=\textwidth]{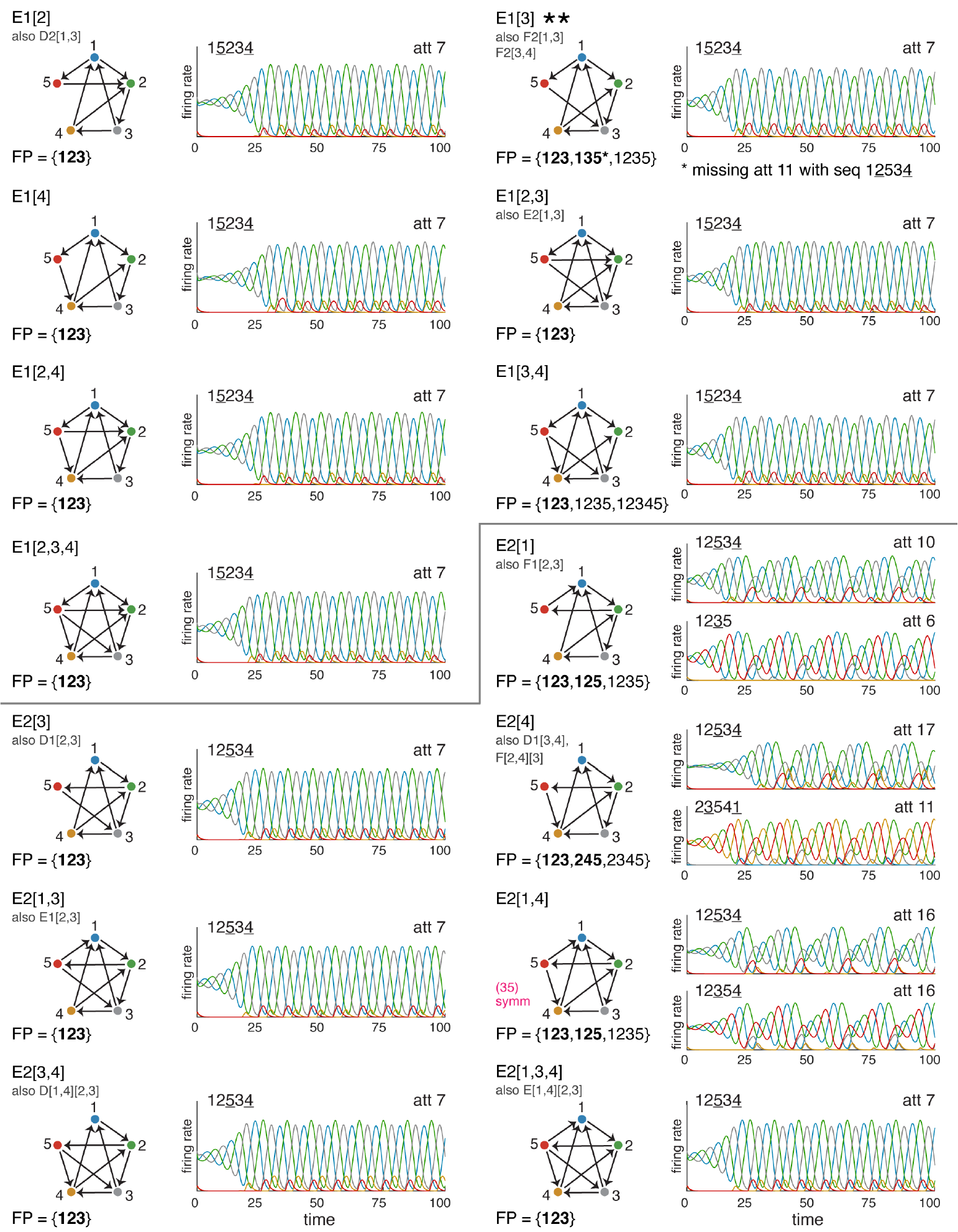}
\end{center}

\newpage
\subsection*{E graphs: E3[$*$] \& E4[$*$]}
\begin{center}
\includegraphics[width=\textwidth]{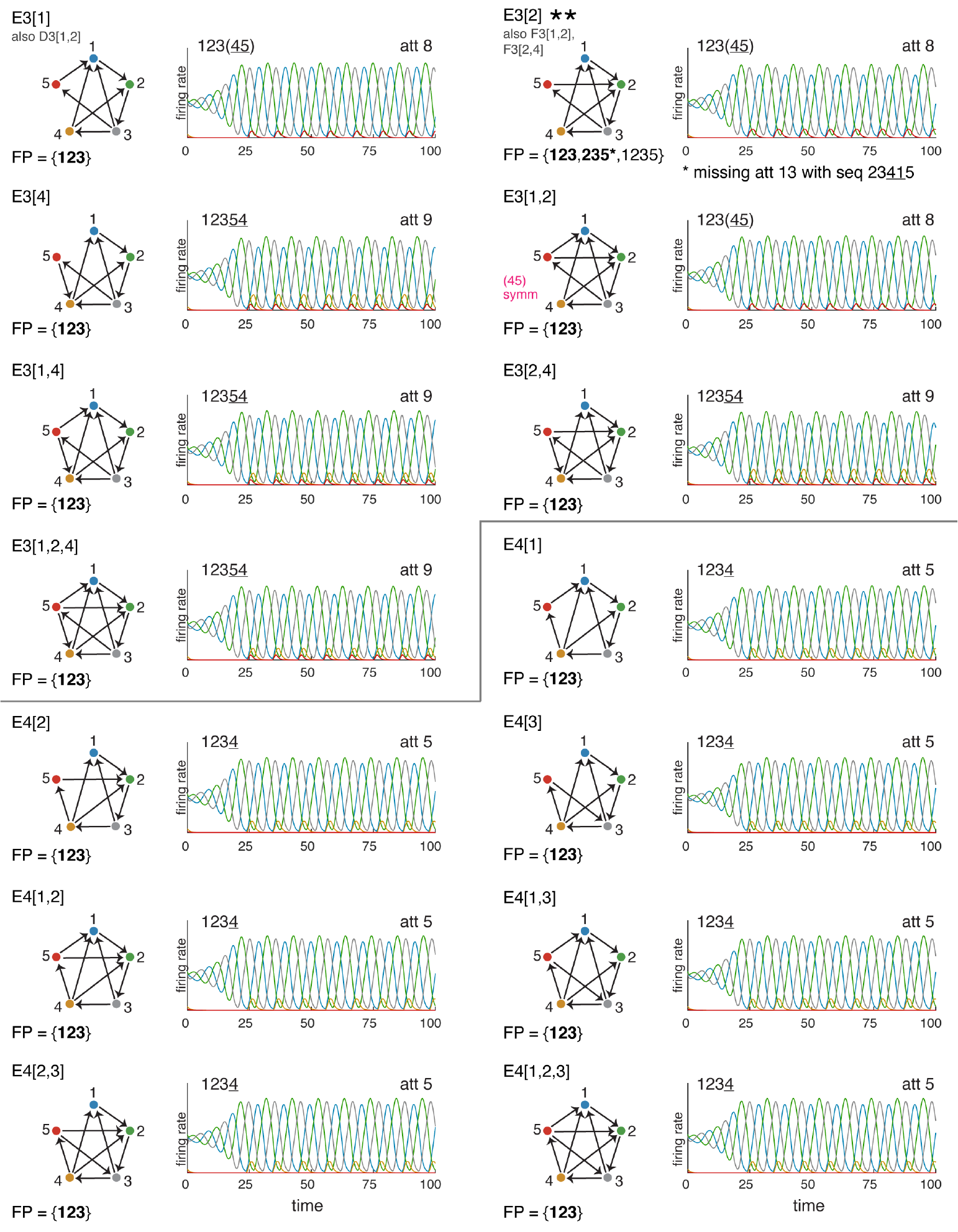}
\end{center}

\newpage
\subsection*{F graphs: F1[$*$] \& F2[$*$]}
\begin{center}
\includegraphics[width=\textwidth]{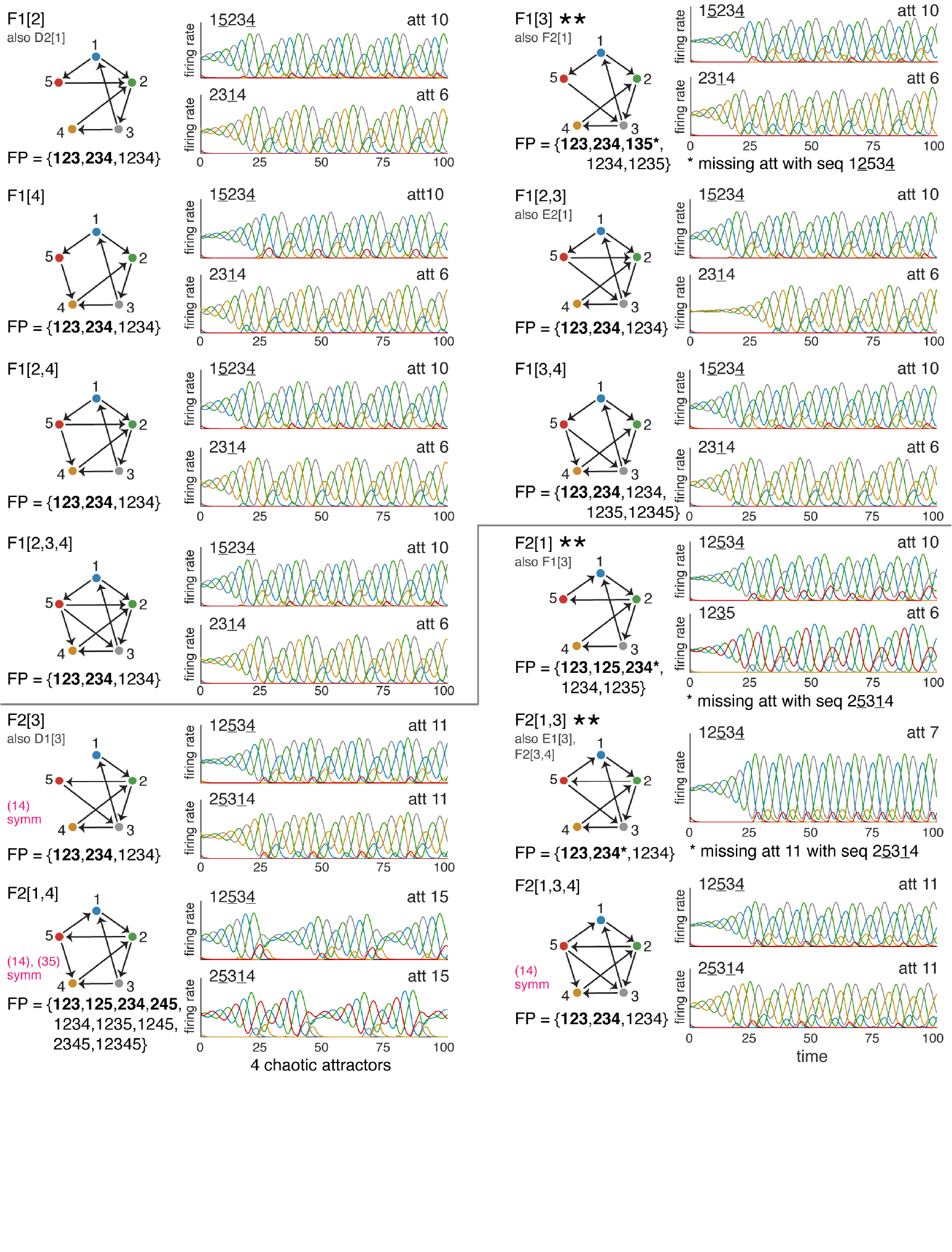}
\end{center}

\newpage
\subsection*{F graphs: F3[$*$]}
\begin{center}
\includegraphics[width=\textwidth]{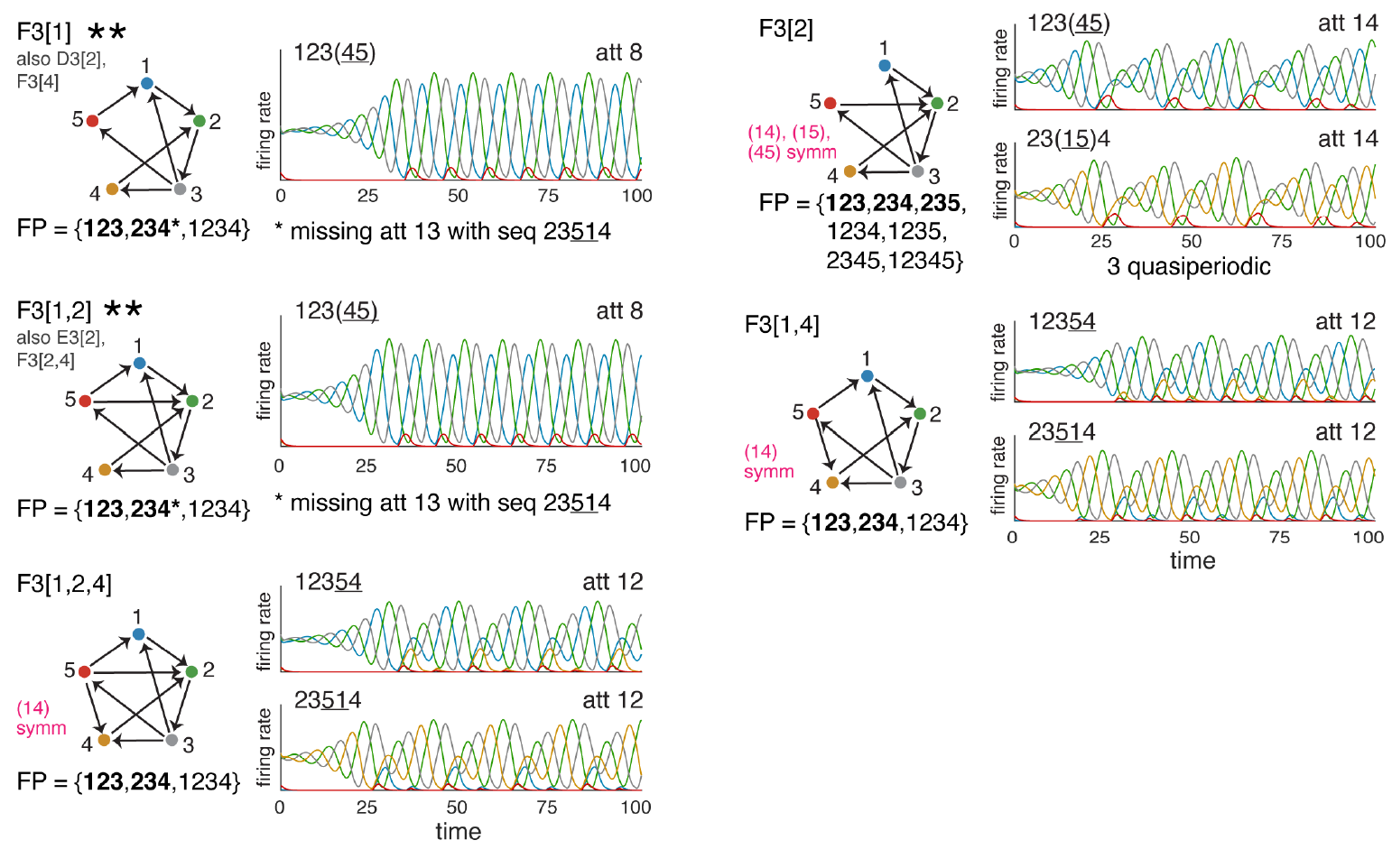}
\end{center}

\newpage
\subsection*{T and S graphs: T4[$*$] \& S1[$*$]}
\begin{center}
\includegraphics[width=\textwidth]{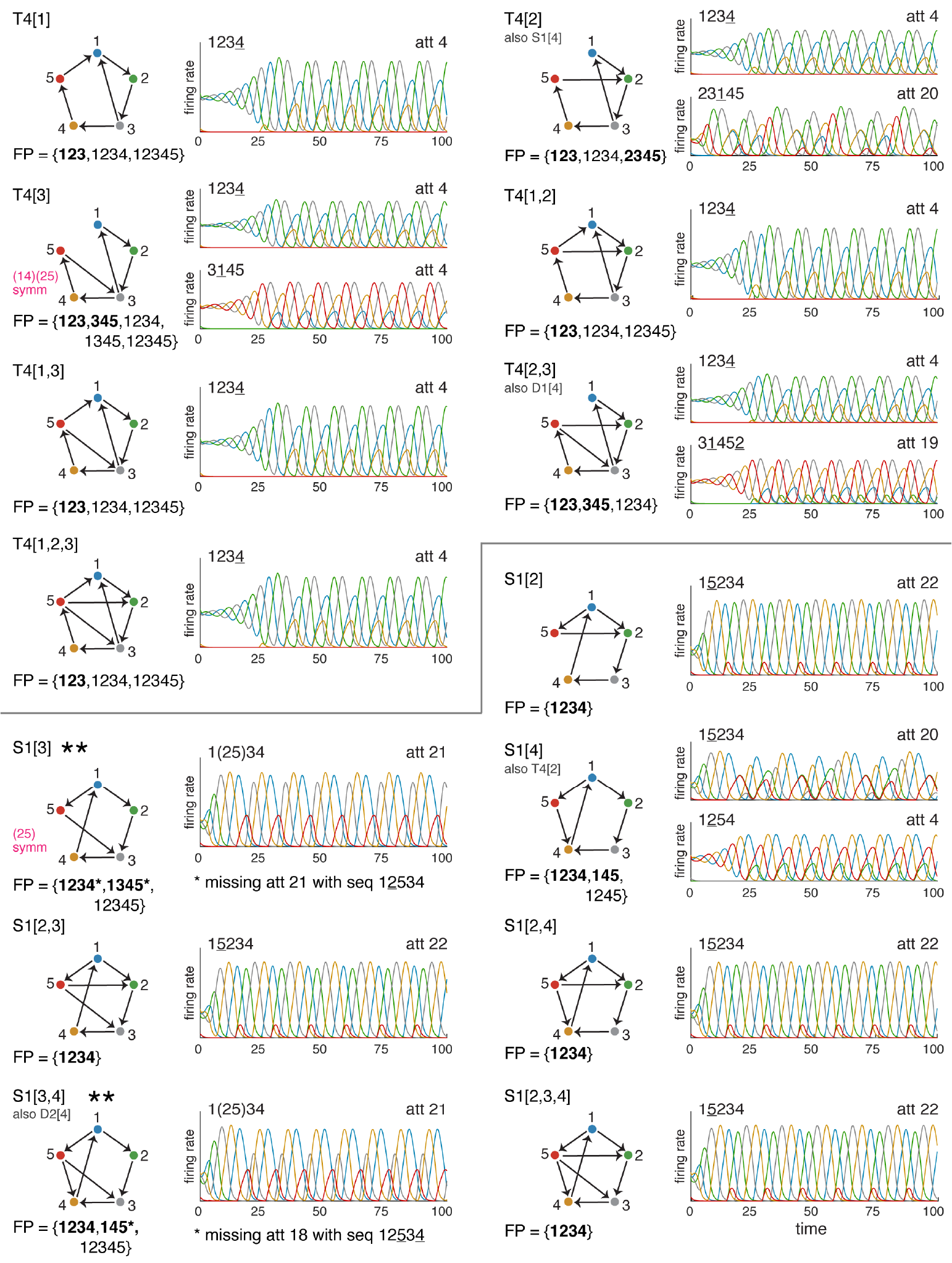}
\end{center}

\newpage
\subsection*{Oriented $n=5$ core motifs}
Recall that the $5$-cycle is the only oriented graph on $n=5$ nodes with no sources or sinks that cannot be constructed from one of the D, E, F, T or S base graphs. It also happens to be a core motif. But there are four additional $n=5$ core motifs, and these can all be constructed from the base graphs. However, they each require two incoming edges to the fifth node, since the added node must eliminate the core fixed point from the cycle in the base graph. Together, they comprise the last three graph families in Table~\ref{table:dictionary}. Note that E[1,2][3,4], drawn here to make apparent the cyclic symmetry, has an additional spurious attractor in the non-standard parameters $\varepsilon = 0.1, \delta = 0.12.$

\begin{center}
\includegraphics[width=\textwidth]{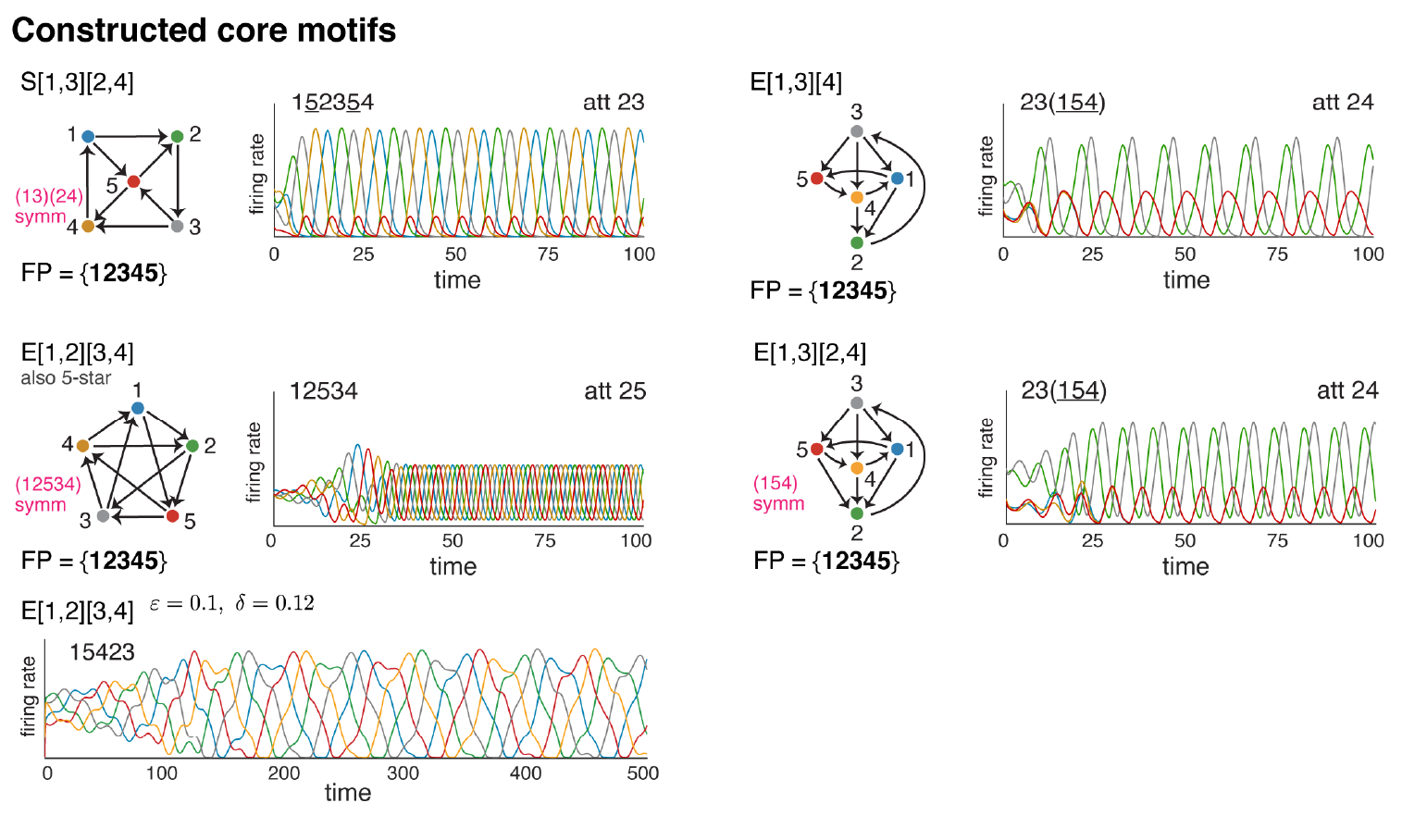}
\end{center}

\newpage
\subsection*{Graphs with ghost attractors, revisited}

Table~\ref{table:n5-count-supp} summarizes the total numbers of core fixed points and attractors for $n=5$ oriented graphs with and without sources. 

\begin{table}[!ht]
\centering
\begin{tabular}{c|c|c|c|c|c}
 graphs & \# graphs & \# core fps & \# attractors & \# ghost atts & \# spurious atts\\
 \hline
 \hline
 with a source & 76 & 87 & 87 & 0 & 0\\
 \hline
with no source & 76 & 104 & 98 & 6 & 0\\
\hline
 total & 152 & 191 & 185 & 6 & 0\\
 \hline
\end{tabular}
\vspace{.15in}
\caption{Core fixed points and attractors for $n=5$ oriented graphs with no sinks. The attractors were found in CTLNs with the standard parameters, $\varepsilon = 0.25, \delta = 0.5.$ }
\label{table:n5-count-supp}
\end{table}

Note that there are 6 core fixed points that did not exhibit a corresponding attractor. Here we see that these so-called ``ghost'' attractors can be recovered as realizable attractors in a different parameter regime. Specifically, if we simulate CTLNs with a significantly higher $\delta$, causing the strong inhibition to be stronger, we end up obtaining a distinct attractor for each of the core fixed points in the problematic graphs. In the figure below, the ``ghost'' attractors are labeled as ``missing in standard parameters." Moreover, by comparing the attractors across graphs in this parameter regime, we were able to identify attractor classes for all but one of the ghosts.

\begin{center}
\includegraphics[width=\textwidth]{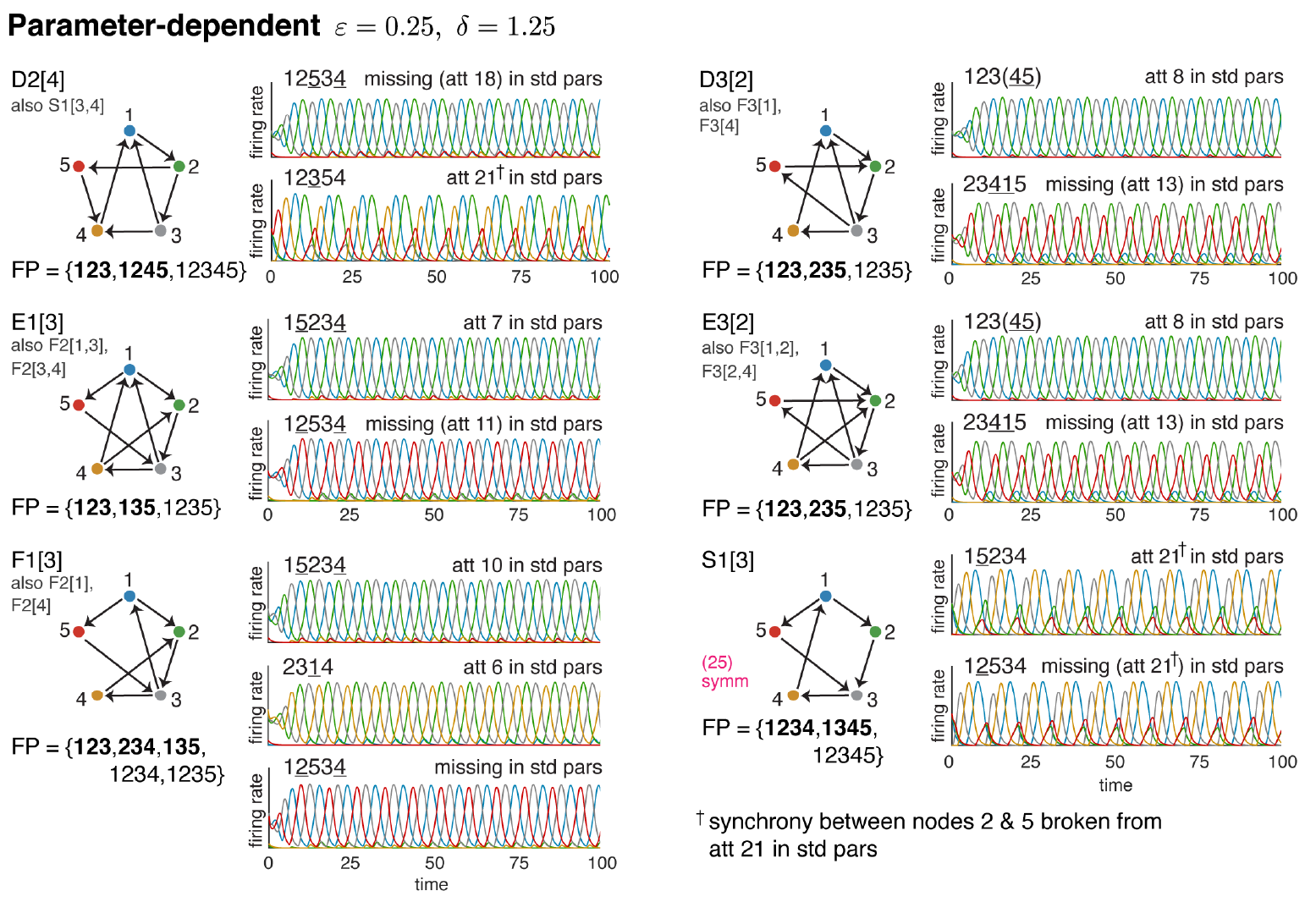}
\end{center}

\newpage

\section{Classification of dynamic attractors for 
$\mathbf{\boldsymbol{\varepsilon} = 0.25, \boldsymbol{\delta} = 0.5}$}

Using the dictionary, we can now collect graphs that exhibit the same attractor into more specific graph families. It turns out that graphs with the same attractor fit nicely together into structural families defined by common and optional graph edges. Table~\ref{table:att-classes-supp} gives the sequence and the graph families corresponding to each of the 25 attractor classes. Note that ``$\sim$'' indicates a forbidden edge. For example, D2[$\sim$3,4,$*$] represents the pair of graphs D2[1,4] and D2[4], which have no edge to node $3$.

\begin{table}[ht]
\centering
\begin{tabular}{l|l|l|c|c|c}
attractor & sequence & graph families & \# graphs & \# attractors & \# ghosts\\
 \hline
 \hline
 att 1 & 123 & 3-cycle + sources & 30 & 30 & 0\\
  \hline
 att 2 & 1234 & 4-cycle + source (aka S0[$*$]) & 5 & 5 & 0\\
  \hline
 att 3 & 12345 & 5-cycle & 1 & 1 & 0\\
  \hline
 att 4 & 123\underline{4} & T4[$*$] & 7 & 8 & 0\\
  \hline
 att 5 & 123\underline{4} & D/E0[$*$] \& D/E4[$*$] & 30+14=44 & 30+14=44 & 0\\
  \hline
 att 6 & 123\underline{4} & F0[$*$] \& F4[$*$] & 11+7=18 & 22+7=29 & 0\\
  \hline
 att 7 & 1\underline{5}23\underline{4} & D1[2,$*$], E1[$*$], \& D/E[1,4][2,$*$] & 4+7+4=15 & 4+7+4=15 & 0\\
  \hline
 att 8 & 123(\underline{45}) &D/E3[$\sim$4,$*$] & 5 & 5 & 0\\
  \hline
 att 9 & 123\underline{54} & D/E3[4,$*$] & 8 & 8 & 0\\
  \hline
 att 10 & 1\underline{5}23\underline{4} & F1[$*$] & 7 & 7 & 0\\
  \hline
 att 11 & 12\underline{5}3\underline{4} & F2[3,$*$] \& F[2,4][3]  & 3+1=4 & 4+1=5 & 1\\
  \hline
 att 12 & 123\underline{54} & F3[1,4,$*$] & 2 & 4 & 0\\
  \hline
 att 13 & 123\underline{54} & F3[$\sim$1,4,$*$] & 2 & 0 & 2\\
  \hline
 att 14 & 123(\underline{45}) & F3[2] & 1 & 3 & 0\\
  \hline
 att 15 & 12\underline{5}3\underline{4} & F2[1,4] & 1 & 4 & 0\\
  \hline
 att 16 & 12\underline{5}3\underline{4} & E2[1,4] & 1 & 2 & 0\\
  \hline
 att 17 & 12\underline{5}3\underline{4} & E2[4] & 1 & 1 & 0\\
  \hline
 att 18 & 12\underline{5}3\underline{4} & D2[$\sim$3,4,$*$] & 2 & 2 & 1\\
  \hline
 att 19 & 1\underline{5}23\underline{4} & D1[4] & 1 & 1 & 0\\
  \hline
 att 20 & 1\underline{5}234 & S1[4] & 1 & 1 & 0\\
  \hline
 att 21 & 1(25)34 & S1[$\sim$2,3,$*$] & 2 & 2 & 1\\
  \hline
 att 22 & 1\underline{5}234 & S1[2,$*$] & 4 & 4 & 0\\
  \hline
 att 23 & 1\underline{5}23\underline{5}4 & S[1,3][2,4] & 1 & 1 & 0\\
  \hline
 att 24 & 23(\underline{154}) & E[1,3][4,$*$] & 2 & 2 & 0\\
  \hline
 att 25 & 12534 & E[1,2][3,4] & 1 & 1 & 0\\
 \hline
\end{tabular}
\vspace{.15in}
\caption{Graph families for attractor classes of $n=5$ oriented graphs with no sinks.}
\label{table:att-classes-supp}
\end{table}

\begin{figure}[!ht]
\begin{center}
\includegraphics[width=.75\textwidth]{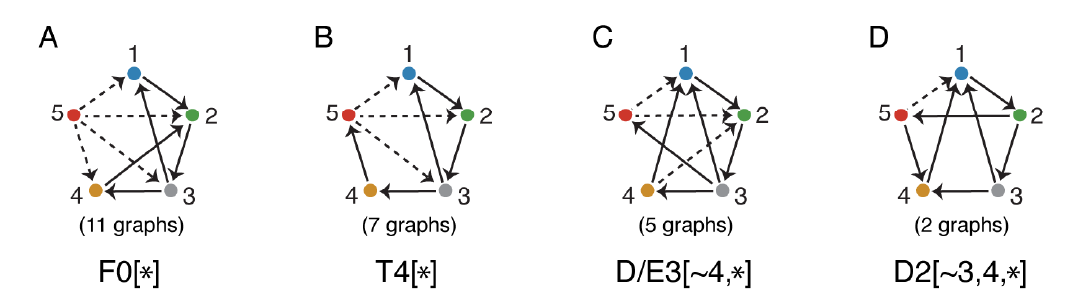}
\caption{\bf Master graphs and their corresponding graph families.}
\label{fig:master-graphs}
\end{center}
\vspace{-.2in}
\end{figure}

In the following pages, graph families for each attractor are depicted graphically, with dashed lines indicating optional edges. Fig~\ref{fig:master-graphs} gives several examples of these {\it master graphs}, and the graph families they represent.  In Fig~\ref{fig:master-graphs}A, all edges from the source node $5$ back to the F graph are optional. Keeping in mind that $5$ must have at least one outgoing edge, and the F graph has a (1,4) symmetry, we count 11 non-isomorphic graphs. In Fig~\ref{fig:master-graphs}A, all $2^3-1 = 7$ combinations of optional edges produce distinct graphs. The master graph in Fig~\ref{fig:master-graphs}C is the union of D3[$\sim$4,$*$] and E3[$\sim$4,$*$]: whether a graph has a D or E base is determined by the presence or absence of the $4 \to 2$ edge. Since $5$ has two optional outgoing edges, there are 3 graphs with base D and 3 with base E in this family. However, the total number of non-isomorphic graphs is 5. This is because D3[1,2] $\cong$ E3[1], as shown in Fig~\ref{fig:naming-graph}. Finally, Fig~\ref{fig:master-graphs}D shows a master graph with only one optional edge. Since $5 \to 4$, both $5 \to 1$ and $5 \not\to 1$ options are allowed, yielding 2 graphs.

For each attractor class, in addition to showing the master graph(s) we also explicitly list all names for graphs that do not have sources, and were thus included in the dictionary. Moreover, we indicate with a superscript if the attractor in question was the primary or secondary attractor for the labeled graph (if a graph is listed without a superscript, it has a single core fixed point and a single attractor). For example, D2[1]$^1$ is listed under att 10, indicating that this attractor corresponds to the first core fixed point for D2[1], which is supported on 123 (see Dictionary). On the other hand, D2[1]$^2$ is listed under att 6, indicating that this attractor corresponds to the second core fixed point, supported on 125. Note that the att 6 and att 10 graph families are most naturally expressed with an F graph base, and D2[1] $\cong$ F1[2] $\cong$ F4[2]. 

It is important to keep in mind that the following classification of dynamic attractors was obtained for the standard parameters. At different parameters, some classes may split and others may merge. However, the prediction from core fixed points remains the same and $\FP(G)$ is invariant. Moreover, the graph families display structure that transcends the choice of parameters. It turns out that all ghost attractors also fit into one of these graph families, with the exception of F1[3]$^3$, corresponding to the third core fixed point supported on 135.

\newpage
\subsection*{Attractor classes att 1--10}
\begin{center}
\includegraphics[width=\textwidth]{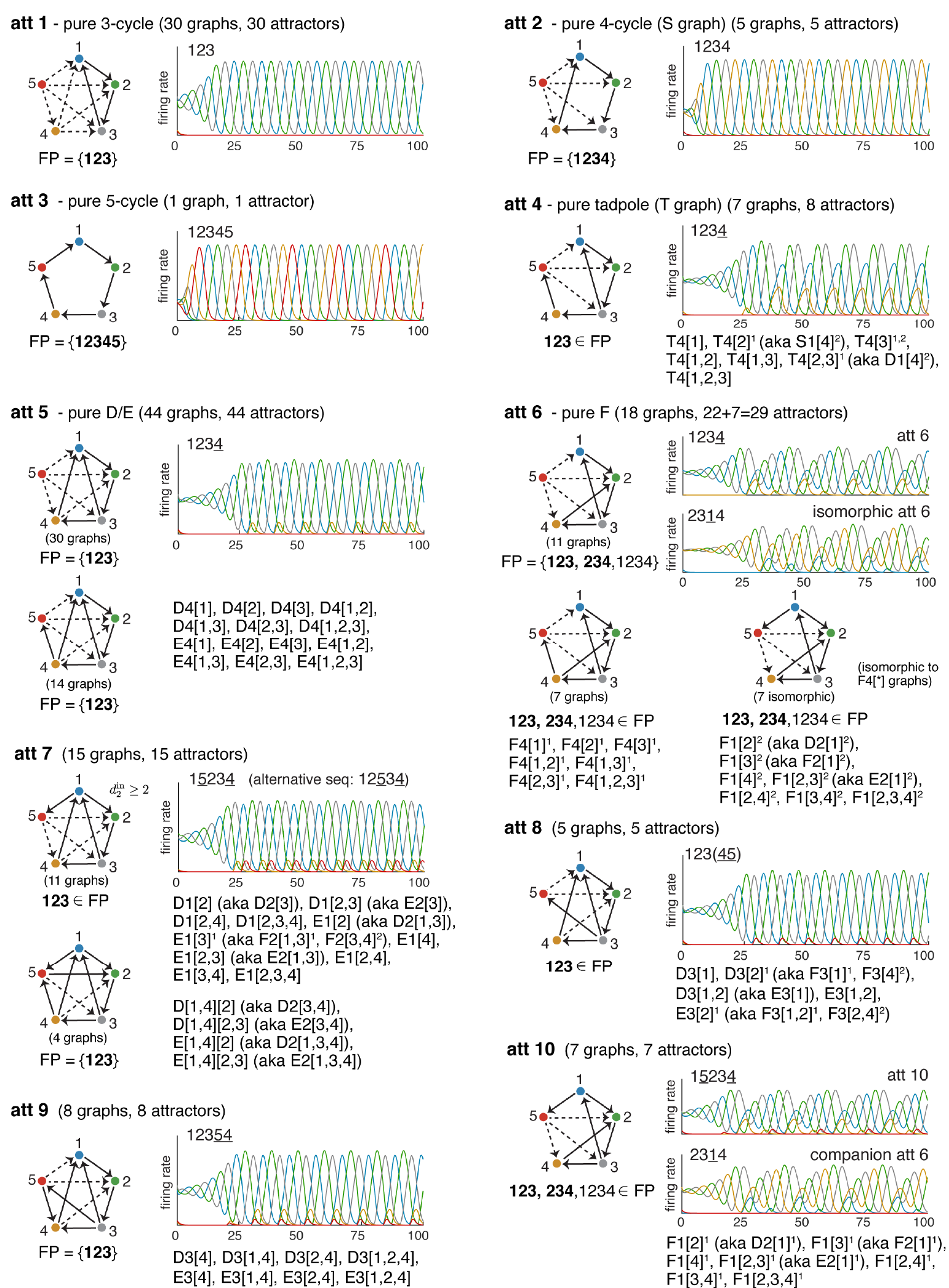}
\end{center}

\newpage
\subsection*{Attractor classes att 11--19}
\begin{center}
\includegraphics[width=\textwidth]{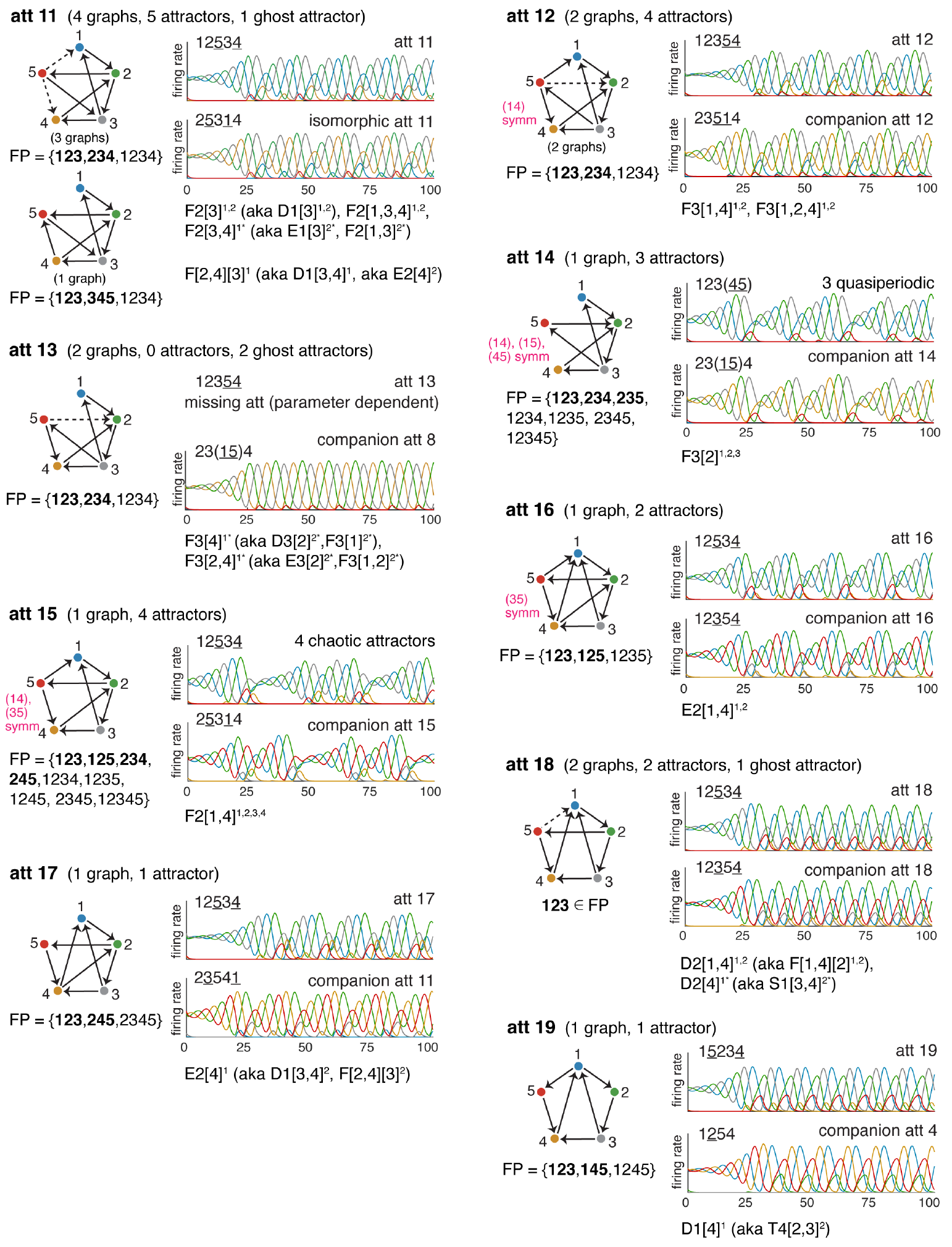}
\end{center}

\newpage
\subsection*{Attractor classes att 20--25}
\begin{center}
\includegraphics[width=\textwidth]{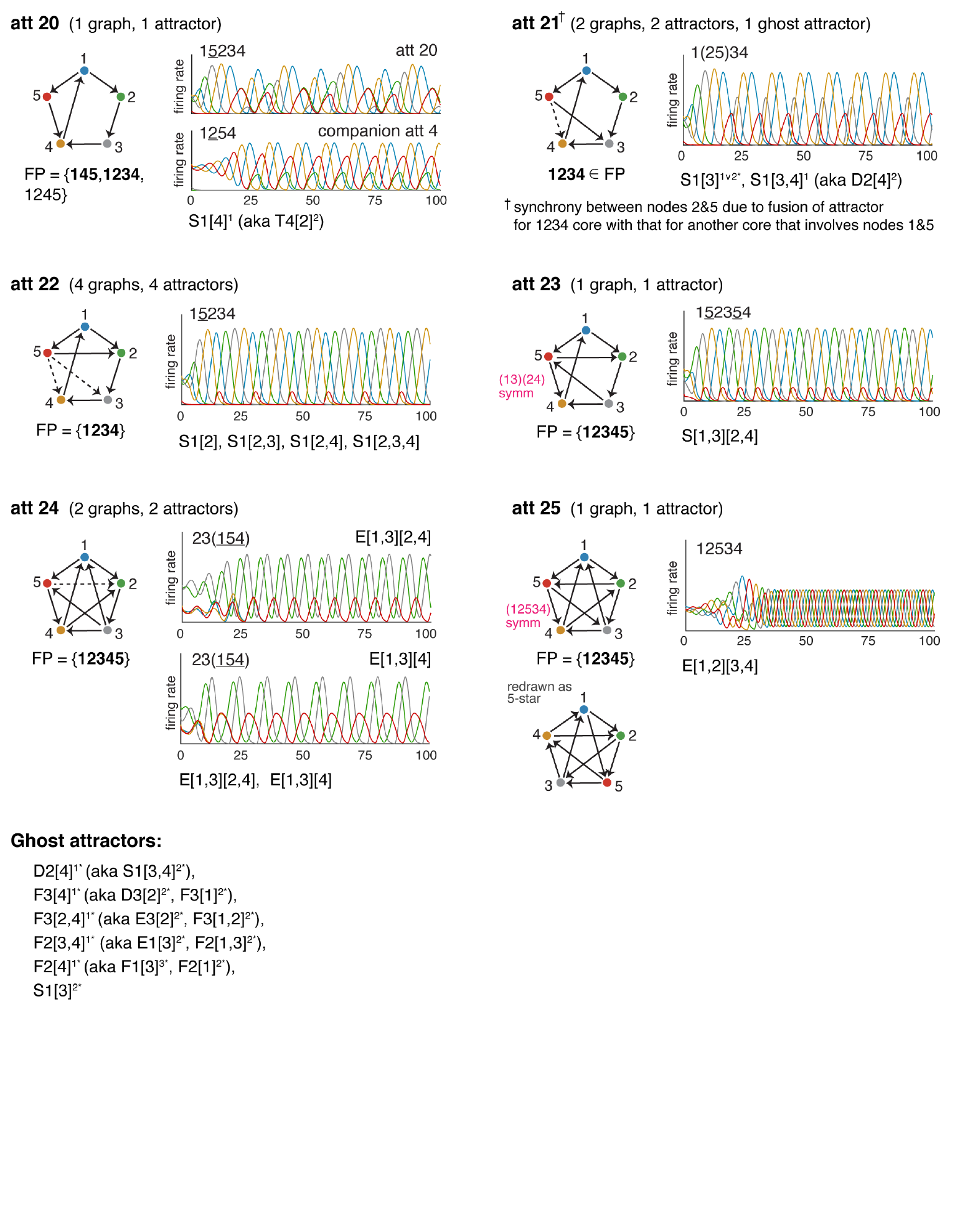}
\end{center}

\bibliographystyle{unsrt}
\bibliography{CTLN-refs}

\end{document}

%% file: packages-v2.tex
\usepackage{amsmath}
\usepackage{amsthm}
\usepackage{amsfonts}
\usepackage{amssymb}
\usepackage{latexsym} 
\usepackage{epsfig}
\usepackage{graphicx}

\usepackage{calc}  
\usepackage[matrix,tips,graph,curve,web]{xy}

\makeatletter

\def\@maketitle{%
  \newpage
  \null
  \let \footnote \thanks
    {\normalfont\sffamily\bfseries\Large\noindent\@title \par}%
    \vskip 1em%
    {\normalfont\sffamily 
        \noindent
        \@author
        \par}
  \par
  \vskip 4em}
\def\@seccntformat#1{\csname the#1\endcsname{.\ }}
\renewcommand\section{\@startsection {section}{1}{\z@}%
                                   {-3.0ex \@plus -1ex \@minus -.2ex}%
                                   {1.5ex \@plus.2ex}%
                                   {\normalfont\large\bfseries}}
\renewcommand\subsection{\@startsection{subsection}{2}{\z@}%
                                     {-2.75ex\@plus -1ex \@minus -.2ex}%
                                     {1.5ex \@plus .2ex}%
                                   {\normalfont\large}}
\def\fnum@figure{\normalfont\footnotesize\figurename~\thefigure}

\renewcommand\tableofcontents{%
    \section*{\contentsname
        \@mkboth{%
           \MakeUppercase\contentsname}{\MakeUppercase\contentsname}}%
    \@starttoc{toc}%
    }
\renewcommand*\l@part[2]{%
  \ifnum \c@tocdepth >-2\relax
    \addpenalty\@secpenalty
    \addvspace{2.25em \@plus\p@}%
    \begingroup
      \setlength\@tempdima{3em}%
      \parindent \z@ \rightskip \@pnumwidth
      \parfillskip -\@pnumwidth
      {\leavevmode
       \large \bfseries #1\hfil \hb@xt@\@pnumwidth{\hss #2}}\par
       \nobreak
       \if@compatibility
         \global\@nobreaktrue
         \everypar{\global\@nobreakfalse\everypar{}}%
      \fi
    \endgroup
  \fi}
\renewcommand*\l@section[2]{%
  \ifnum \c@tocdepth >\z@
    \addpenalty\@secpenalty
    \addvspace{1.0em \@plus\p@}%
    \setlength\@tempdima{1.5em}%
    \begingroup
      \parindent \z@ \rightskip \@pnumwidth
      \parfillskip -\@pnumwidth
      \leavevmode \sffamily\bfseries
      \advance\leftskip\@tempdima
      \hskip -\leftskip
      #1\nobreak\hfil \nobreak\hb@xt@\@pnumwidth{\hss #2}\par
    \endgroup
  \fi}
\renewcommand*\l@subsection{\sffamily\@dottedtocline{2}{1.5em}{2.3em}}
\renewcommand*\l@subsubsection{\@dottedtocline{3}{3.8em}{3.2em}}
\renewcommand*\l@paragraph{\@dottedtocline{4}{7.0em}{4.1em}}
\renewcommand*\l@subparagraph{\@dottedtocline{5}{10em}{5em}}
\makeatother


\theoremstyle{plain}
\newtheorem{theorem}{Theorem}

\newtheorem{lemma}[theorem]{Lemma}

\theoremstyle{definition}
\newtheorem{definition}[theorem]{Definition}

  {\begin{list}{}%
    {%
    \settowidth{\labelwidth}{#1}%
    \setlength{\itemindent}{0pt}%
    \setlength{\labelsep}{1em}%
    \setlength{\leftmargin}{\labelwidth+\parindent+\labelsep}%
    \setlength{\itemsep}{0pt}%
    \setlength{\parsep}{.6ex}}}%
  {\end{list}}

  {\begin{list}{}%
    {%
    \settowidth{\labelwidth}{#1}%
    \setlength{\itemindent}{0pt}%
    \setlength{\labelsep}{1em}%
    \setlength{\leftmargin}{\labelwidth+\labelsep}%
    \setlength{\itemsep}{0pt}%
    \setlength{\parsep}{.6ex}}}%
  {\end{list}}

\makeatletter
\newenvironment{subequations*}{
  \begingroup 
  \let\protect\@nx
  \edef\@tempa{\def\@nx\theparentequation{\theequation}}%
  \@xp\endgroup\@tempa
  \setcounter{parentequation}{\value{equation}}%
  \setcounter{equation}{0}%
  \def\theequation{\theparentequation\alph{equation}}%
  \ignorespaces
}{%
  \setcounter{equation}{\value{parentequation}}%
  \global\@ignoretrue
}

\def\d/{/\mspace{-6.0mu}/}